\documentclass[12pt]{article}
\usepackage{epsfig, amssymb, amsmath}
\usepackage{graphicx,epsfig}
\usepackage{color}
\newcommand{\fixme}[2]{\textcolor{black}{{#1}}}
\newcommand{\fixmeag}[2]{\textcolor{black}{{#1}}}
\begin{document}
\begin{titlepage}
\thispagestyle{empty}
\begin{flushright}
\end{flushright}\textbf{}

\bigskip

\begin{center}
\noindent{\Large \textbf
{Partially massless theory as a quantum gravity candidate}}\\

\vspace{2cm} 
\noindent{Lunchakorn Tannukij${}^{a}$\footnote{e-mail:l\_tannukij@hotmail.com} 
 and Jae-Hyuk Oh${}^{b}$\footnote{e-mail:jaehyukoh@hanyang.ac.kr}}
\vspace{1cm}

{\it
Department of Physics, Hanyang University \\
 Seoul 133-891, Korea${}^{a}$\\
}
\end{center}

\vspace{0.3cm}
\begin{abstract}
We study partially massless gravity theory(PM gravity theory) and suggest an alternative way to add higher order interaction vertices to the theory. Rather than introducing self interaction vertices of the gravitational fields to the partially massless gravity action, we consider interactions with matter fields, since it is well known that addition of the self interaction terms necessarily breaks the $U(1)$ gauge symmetry that PM gravity theory enjoys. 
For the coupling with matter fields, we consider two different types of interaction vertices. The first one is given by an interaction Lagrangian density, $\mathcal L_{int}\sim h_{\mu\nu}T^{\mu\nu}$, where $h_{\mu\nu}$ is the PM gravity field and $T^{\mu\nu}$ is the stress-energy tensor of the matter fields. To retain the $U(1)$ gauge symmetry, the matter fields also transform accordingly and it turns out that the transform must be non-local in this case. The second type interaction is obtained by employing a gauge covariant derivative with the PM gravity field, where the PM gravity fields play a role of a gauge connection canceling the phase shift of the matter fields.
We also study the actions and the equations of motion of the partially massless gravity fields. As expected, it shows 4 unitary degrees of freedom. 2 of them are traceless tensor modes and they are light like fields. The other 2 are transverse vector modes and their dispersion relation changes as background space time (de Sitter) evolutes. In the very early time, they are light like but in the very late time, their velocity become a half of speed of light. The vector mode dispersion relation shows momentum dependent behavior. In fact, the higher(lower) frequency modes show the faster(slower) velocity. We call this effect ``conformal(or de Sitter) prism". We suggest their quantization, compute Hamiltonians to present their exited quanta and construct their free propagators.
\end{abstract}
\end{titlepage}

\newpage

\tableofcontents
\section{Introduction}
Einstein gravity theory is a great success to explain how the matters(more precisely their energy and momenta) affect their space time that they are sitting in. The Mercury's perihelion advance in our solar system, the gravitational lensing  and so on are the unavoidable facts that we can observe from the sky and they show that Einstein's gravity must be correct. Moreover, recently, the gravitational waves are detected, which are expected by a weak field expansion of the Einstein equations(The linearized Einstein gravity). Somehow, there is no reason that we do not accept the Einstein gravity as a well designed classical theory which perfectly describe the gravitational effects in our world.

However, the wave-particle duality in quantum mechanics leads this well designed theory to puzzles. Does graviton exist? Quantum field theory provides reasonable processes which quantize the fields and the results are appearance of quanta with definite energy and momenta, which we call ``particle''. These particles interplay with themselves or the other particles and give probabilities or interaction strengths that the particles effectively interact one another. If the graviton exists, then does it provide an effective gravitational theory? Is this theory the Einstein gravity or another? Which theory should be a right candidate for the quantum gravity? So far, no one can answer these questions.

The first try to quantize the gravitational theory is to quantize the Fi\fixmeag{er}{}z-Pauli action\fixmeag{.}{with/without the gravitational mass}
In \cite{Fierz:1939ix}, Fierz and Pauli discuss what kind of spin 2 field action gives transverseness and relativistic dispersion relation to the fields in the level of its equation of motion. It turns out that Fierz-Pauli action\fixmeag{, even }{with/}without its mass term, \fixmeag{also known as linearized Einstein-Hilbert action,}{} in the flat background does. 
In fact, the gravitational fields from the quadratic truncation  of the weak field expansion of the Einstein-Hilbert action are transverse and show null propagation. 

Another issue to quantize the gravitational fields is the ghost problem. In \cite{VanNieuwenhuizen:1973fi}, the authors
discuss the condition that the Hamiltonian of the spin-2 fields  becomes positive definite. The Fi\fixmeag{er}{re}z Pauli action\fixmeag{, thanks to the well-chosen mass term,}{}  is the unique action without ghost and tachyon states \cite{Fierz:1939ix}. \fixmeag{}{with/without mass terms.}

The linearized gravity looks perfect as a candidate for quantum gravity, but it cause\fixmeag{s}{} some problems when the interactions are taken into account\cite{Veltman:1975vx}. For example, the \fixmeag{linearized gravity (the $m=0$ Fierz-Pauli action)}{massless Fierz Pauli action} enjoys diffeomorphism invariance and it fixes the self interactions and the interactions with matters. It turns out that it is nonrenormalizable from 2 loop diagrams. 
 
To circumvent these difficulties, people try many of the other types of quantum gravities. One of the directions as such is a study on quantum gravities in de Sitter background. 
In fact, we are living in de Sitter space. The cosmological constant of our universe is very small($<(10^{-9}{\rm MeV})^4$) and positive and \fixmeag{believed to be responsible for the expansion of our universe}{our universe is expanding}. Followed by these facts, there have been many of tries to get gravitational field propagator in de Sitter space\cite{Christensen:1979iy,Allen:1986tt} of the linearized Einstein gravity.


Partially massless gravity is a massive gravity theory in de Sitter space, defined at a special point of the parameter space of the gravit\fixmeag{on}{ational} mass ``$m$'' and the cosmological constant ``$\Lambda$''. 
When one adds the mass term to the \fixmeag{linearized Einstein-Hilbert}{Fierz-Pauli} action, \fixmeag{hence the Fierz-Pauli action}{}, it breaks the diffeomorphism invariance of the theory. However, when   
the mass $m^2=\frac{2\Lambda}{3}$ in the de Sitter background, there is a noncompact $U(1)$ symmetry enhanced and this play\fixmeag{s}{} interesting role.
\footnote
{There are various directions of studies on the PM gravity theories. Supersymmetric PM gravity\cite{Garcia-Saenz:2018wnw}, PM gravity in various spacetimes\cite{Bernard:2017tcg}, solitonic solutions in PM gravity theory\cite{Hinterbichler:2015nua}, PM gravity in 3-dimension\cite{Alexandrov:2014oda} are considered. There are several consistency and validity checks for PM gravity  and/or massive gravity theories, too. For examples,  Cosmic tests for massive gravity\cite{Enander:2015dpn}, validity checks by employing the method of characteristics\cite{Deser:2014fta} and stability checks for black hole solutions in massive gravity theories\cite{Brito:2013yxa} are there.
}

Massless \fixmeag{$(m=0)$}{} Fierz-Pauli action has 2 degrees of freedom, which are traceless transverse tensors. This is due to the general covariance or diffeomorphism invariance. However, the \fixmeag{generic}{massive} Fierz-Pauli action has 5 degrees of freedom, which contains 2 transverse vector \fixmeag{degrees of freedom (dofs)}{dof} and 1 scalar dof together with the 2 tensor dof\fixmeag{s}{}.
\fixmeag{Normally, if one considers a massive gravity theory of general mass term (not the Fierz-Pauli one), there will be another (sixth) scalar degree of freedom which}{Normally, this additional scalar degree of freedom} causes negative energy states since the action for this \fixmeag{sixth scalar mode}{} is not a two derivative one but contains higher order time derivatives\cite{Chen:2012au}. 

In fact, the partially massless condition, $m^2=\frac{2\Lambda}{3}$ eliminates the \fixmeag{(fifth)}{} scalar degree of freedom. When the $U(1)$ gauge symmetry is enhanced, the scalar mode is no longer a real degree but it becomes gauge degree of freedom. One can eliminate this by a gauge choice. Therefore, it turns out that partially massless gravity theory contains 4 healthy degrees of freedom.

However, nonlinear extensions of the partially massless gravity theory faces very serious problems\cite{deRham:2013wv}. One of the problems is the following. If one adds nonlinear(multi-point interaction) terms to the PM action as $\sim\int \sqrt{g}{\ }d^4x {\ }h_{\mu\nu}^n(x)$, where $n>2$, the $U(1)$ gauge symmetry is manifestly broken.

To understand why it is so, let us look at quantum electrodynamics(QED). 
QED contains massless vector fields $\mathcal A_\mu$ which enjoy $U(1)$ gauge symmetry. If one adds either their mass terms 
or multi-point interaction terms(without gauge covariant derivatives in them) $\sim \int\sqrt{g}{\ }d^4x  [\mathcal A_\mu(x) \mathcal A^\mu(x)]^n $, where $n\geq1$, the gauge symmetry is broken manifestly. In fact, there is no self-interactions between the gauge fields in QED.
One may add higher order interaction terms containing their field strength.
However, these provide higher order time derivatives to the theory. In general the Hamiltonian(or energy)of such a theory is not bounded from below, namely one cannot find vacuum states of the theory\cite{Chen:2012au}. It is \fixmeag{often}{open} called ghost problem, where particles with negative energy states appear. 
The situation in PM gravity theory is more or less like this. In the many of the literatures, there are attempts to add interaction terms without derivatives in them but it turns out that they are all failed \fixmeag{\cite{deRham:2013wv,Deser:2013uy}}.

Even though there are such restrictions, the $U(1)$ gauge fields show interactions in QED. Rather than adding their self interaction terms, the gauge fields interact with fermion fields. As long as the fermion current is conserved, the gauge symmetry is retained. However, the fermion current conservation is hold only for the on-shell fields since the Noether charge is conserved \fixmeag{up to}{upto} the  equation of motion.  To request the $U(1)$ gauge symmetry for the off-shell fields, the fermion fields need to transform properly under the gauge transformation. In fact, the fermion phase changes under it.

Partially massless theories enjoy the similar property.
By observing the similarity between QED and PM gravity theories, one probably realize that developing gravitational self-interaction must violate the $U(1)$ gauge symmetry that PM gravity has. Rather than this way, one needs to construct interactions between the PM gravitational fields and matter fields.

More precisely, the $U(1)$ gauge symmetry transform of the PM gravitational fields is given by
\begin{equation}
\delta h_{\mu\nu}=\nabla_\mu\nabla_\nu \phi + \frac{\Lambda}{3} g_{\mu\nu}\phi, \label{PMsym}
\end{equation}
where $\nabla$ is the covariant derivative with the de Sitter metric and where the $\phi$ is gauge parameter. When the PM gravitational fields interact with matter fields, the transform\fixmeag{ation}{} becomes symmetry only when the matter current is conserved as
\begin{equation}
\label{current-con}
\nabla_\mu\nabla_\nu T^{\mu\nu} + \frac{\Lambda}{3} T^{\mu}_{\mu}=0. 
\end{equation}
However, the matter current is conserved only when they are on-shell as discussed. Therefore one needs to develop an appropriate transformation of the matter fields so that this becomes a symmetry without requesting their current conservation.

The first issue that we raise in this note is a development of the matter field transformations, where
we consider two types of PM gravity couplings with matters. The first type is characterized by an interaction Lagrangian density being given by $\mathcal L_{int}\sim h_{\mu\nu}T^{\mu\nu}$, where $T^{\mu\nu}$ is the stress energy tensor of the matter field. This comes from the classical notion of general relativity, where the gravitational fields interact with (or space time curvature is given by) energy and momentum of matters.
We show that the matter field transformation is nonlocal one in this case\footnote{Electric-magnetic duality transformation is nonlocal transformation but it is global symmetry\cite{Deser1,Moon:2014gaa,Lee:2018bud,Hinterbichler:2014xga}}.
We can see this due to the fact that we confront a contradiction if we suppose that both of the $\phi$ and the $\zeta_I$ are local parameters where the $\zeta_I$ is the transformation parameter of the matter fields.
For the PM transformation to be a symmetry, a certain  relation between $\zeta_I$ and $\phi$ should come into existence, namely $\zeta_I=\zeta_I(\phi)$. The relation has a fractional form and in general the denominator cannot be removed by canceling the common factors of the numerator and the denominator. The denominator contains certain amount of derivatives which means the transformation does have integrations in it.


For concreteness, we concentrate on the massless scalar and electromagnetic fields($U(1)$ vector fields)\footnote{In \cite{Deser:2006zx}, the authors discuss an interaction between partially massless spin 2 and spin 1 fields. However, the spin 1 fields are not the usual electromagentic fields.}. 
The result is listed in the section \ref{Gauge symmetry on the matter fields}. In section \ref{Gauge symmetry on the matter fields}, we show that in general the matter transformation is nonlocal. After this, we develop a transformation of the matter fields together with the partially massless symmetry transform\fixmeag{ation}{} which makes the full action invariant. 

However, there is several pieces of criticism on the non locality of the transformation. First, non locality may cause a-causality and/or non-unitarity in the theory. These have become reasons of obstruction for non linear extension of PM gravity theory\cite{Joung:2014aba,Apolo:2016ort,Garcia-Saenz:2015mqi}\footnote{There is partial success on this issue. In \cite{Boulanger:2020bah}, the authors construct interactions between partially massless spin 2 fields and Abelian 1-form gauge fields. The Abelian vector fields that they introduced are Stueckelberg fields compensating the gauge transformation from the partially massless spin 2 fields. In \cite{Garcia-Saenz:2014cwa}, the authors consider a consistent condition for the partially massless spin 2 symmetry such that its two successive transformations still respect the symmetry. Under such a condition, they find a non-linear partially massless spin 2 transformation which reduces to the original symmetry transformation when the gauge parameter in the non-linear one is treated as an infinitesimal parameter. They also report that no consistent Lagrangian density upto 2 derivative level is not found but still the possibility of existence of a Lagrangian density in higher order in derivatives is open.}. Secondly, if the non-local transformation is allowed, then the gauge symmetry and the corresponding Noether process
are trivially achieved. Allowing Non-locality is not as much restrictive as local-transformation. The locality gives the genuine and strong constraints on the gauge theory and the theory becomes rather refined. 

To avoid the non locality, we suggest the second type interaction based on a local transform, which is constructed by employing a gauge covariant derivative. We consider an object, $\Gamma^{(I)}_\mu(h_{\alpha\beta})$, of which transformation is given by $\delta \Gamma^{(I)}_\mu(h_{\alpha\beta}) \sim \partial_\mu \beta$ under the PM gravity transform, where $I=1,2,...$, and the $\beta$ is a certain scalar quantity being linear in $\phi$. We utilize these objects as gauge connections such that we promote any derivatives appearing in the matter fields action to $D^{(I)}_\mu \equiv \partial_\mu +i\kappa_5^{I-1}\Gamma^{(I)}_\mu$. The larger index, $_I$, the higher order in derivative and in coupling constant $\kappa_5$ in $\Gamma^{(I)}$. The minimal(the lowest derivative) object is $\Gamma^{(1)\mu}\equiv \nabla_\nu h^{\mu\nu}$.
As examples, we suggest fermion and complex scalar matters interacting with PM gravity fields. 

In the section \ref{Solutions of the equations of motion and their quantization}, we construct quantizations of the (free) traceless tensor modes and the transverse vector modes. The traceless tensor modes show null propagation and its quantization process is pretty much standard. However, the
transverse vector modes are rather peculiar in the following senses. First, it shows a\fixmeag{n}{} unusual dispersion relation and it changes as the Universe evolves.
Moreover, the dispersion relation tells us that the vector particle with the higher momentum travels faster. We call this effect ``conformal prism''. 
We construct the quantization of the vector fields and it turns out that the quantization scheme depends on the age of the Universe. We note that in \cite{Deser:2001xr}, the partially massless particles all show null propagation.
This is true in a sense that as we will show, the vector and tensor mode enjoy time rescaling invariance and so in some frame of time they show null propagations.
However, we need \fixmeag{to}{} choose our gauge of such a symmetry to fix the time since we need to define our vacuum uniquely. When we choose the conformal time, the traceless tensor mode travels in the speed of light but the transverse vector mode does not. For the detailed discussion of the quantization, look at section \ref{Solutions of the equations of motion and their quantization}.

\section{A short review of partially massless gravity}
In usual quantum field theories, one may evaluate free propagators with a quadratic action(free fields action) before considering their interactions with the other fields(or themselves). However, the gravitational field\fixmeag{}{s} provides infinite number of the self interaction terms, 
when one considers a weak field expansion of Einstein-Hilbert action. 
Maybe it is more likely to consider the Einstein-Hilbert action as a quantum effective theory rather than to treat that as a genuine quantum field theory acton. 

However, the Einstein-Hilbert action provides a good motivation for the free theory of gravitation. 
In fact, there are many of literatures studying the quadratic truncation of the weak expansion of the Einstein-Hilbert action (together) with/without the gravit\fixmeag{on}{ational} mass term. One of the main issue that the authors discuss is how to get the ghost free Lagrangian density.

One of the candidates showing ghost-free two point correlators is the partially massless gravity theory, which provides healthy 4  \fixmeag{dofs}{dgrees of freedom(dof)}, which are 2 tensor dof\fixmeag{s}{} and 2 vector dof\fixmeag{s}{}.
The partially massless gravity theory is based on the Fi\fixmeag{er}{re}z-Pauli action with a certain mass term and their interaction with matter fields.
We consider the following action,
\begin{eqnarray}
S&=&\int d^4 x \sqrt{\fixmeag{-}{} g}\left[  -\frac{1}{2}\nabla_\lambda h_{\mu\nu} \nabla^\lambda h^{\mu\nu} + \nabla_\lambda h_{\mu\nu} \nabla^\nu h^{\mu\lambda}
-\nabla_\mu h \nabla_\nu h^{\mu\nu} +\frac{1}{2}\nabla_\mu h \nabla^\mu h   \right. \\ \nonumber
&+&\left.\Lambda\left( h^{\mu\nu}h_{\mu\nu}-\frac{1}{2}h^2 \right)-\frac{1}{2}m^2(h_{\mu\nu}h^{\mu\nu}-h^2)-\frac{1}{4}\mathcal F_{\mu\nu}\mathcal F^{\mu\nu}+h_{\mu\nu}T^{\mu\nu}\right],
\end{eqnarray}
where the background geometry is maximally symmetric and in fact it is given by de Sitter space satisfying
\begin{equation}
 R_{\mu\nu\rho\sigma}=\frac{\Lambda}{3}( g_{\mu{\rho}} g_{{\nu}\sigma}- g_{\mu\sigma} g_{\nu\rho}),
\end{equation}
and the $\Lambda$ is the cosmological constant being positive.
The de Sitter space metric we will use is given by
\begin{equation}
ds^2=-d {\rm t}^2  + e^{2\sqrt{\frac{\Lambda}{3}}{\rm t}}d\vec x^2_3,
\end{equation}
where $d\vec x^2_3=dx^2+dy^2+dz^2$. By defining the conformal time as
$dt=d{\rm t}e^{-\sqrt{\frac{\Lambda}{3}}{\rm t}}$, the metric is given by 
\begin{equation}
\label{conformal-time-metric}
ds^2=\frac{3}{\Lambda t^2}(-dt^2+d\vec x^2_3 ),
\end{equation}
where $-\infty<t<0$.
\footnote{
One may also use a static coordinate as
\begin{equation}
ds^2=-\left(1-\frac{\Lambda r^2}{3}\right)dt^2+{\left(1-\frac{\Lambda r^2}{3}\right)^{-1}}{dr^2}+r^2d\Omega^2_2,
\end{equation}
but we will use (\ref{conformal-time-metric}) in this note.
}

One of the good properties of this theory is that when $m^2=\frac{2\Lambda}{3}$, a new (noncompact) $U(1)$ gauge symmetry is enhanced and the action is invariant under the symmetry transform\fixmeag{ation}{} as \fixmeag{follows,}{}
\begin{equation}
\delta h_{\mu\nu}=\nabla_\mu\nabla_\nu \phi + \frac{\Lambda}{3} g_{\mu\nu}\phi
\end{equation}
under one condition that the stress energy tensor satisfies
\begin{equation}
\label{extra-u1-condition}
\nabla_\mu\nabla_\nu T^{\mu\nu} + \frac{\Lambda}{3} T^\mu_{\ \mu}=0
\end{equation}




\section{Gauge symmetry on the matter fields}
\label{Gauge symmetry on the matter fields}
\subsection{$\mathcal L_{int}\sim h_{\mu\nu}T^{\mu\nu}$-type interaction}
The partially massless symmetry is a certain combination of the diffeomorphism and Weyl transformation. In \fixmeag{a generic}{} massive gravity theory, none of them is\fixmeag{}{not} symmetry. However,
for the \fixmeag{special}{spatial} value of the gravitational mass, this becomes symmetry. We learn from this that Weyl transformation has a role for the symmetry. 

If PM gravity theory couples matters, the symmetry is still retained only when the conservation equation of the energy momentum tensor is satisfied.
The energy momentum conservation is hold only for the on-shell fields. Therefore, if one wants to construct a quantum theory being coupled to matters, the matter fields \fixmeag{must}{} transform in appropriate way under such transformation. It turns out that the transformation rule is nonlinear in the matter and PM gravitational field even at the level of infinitesimal transformation. 

There are some of issues to discuss for the construction of the symmetry transform\fixmeag{ation}{}. Firstly, we discuss the nonlocality of it. Suppose the matter fields enjoy a local transform\fixmeag{ation}{} as $\Phi_I(x^\mu)\rightarrow \Phi_I(x^\mu)+\zeta_I(x^\mu)$, where the index $I$ denotes any indices that the matter fields carry.  The full action $S$ is given by
\begin{equation}
S=S_{PM}+S_{M}\equiv S_{PM}+\int \sqrt{-g}d^4x h_{\mu\nu}(x^\mu)T^{\mu\nu}(x^\mu) +\int\sqrt{-g}d^4x \mathcal L(\Phi_I(x^\mu))
\end{equation}
We want that the full action, $S$ is invariant under $(h_{\mu\nu}(x^\mu),\Phi_I(x^\mu))\rightarrow(h^\prime_{\mu\nu}(x^\mu),\Phi_I^\prime(x^\mu))$.

Consequently, the variation of the full action is given by
\begin{equation}
0=\int \sqrt{-g}d^4x \left[h_{\mu\nu}(x^\mu)\Delta T^{\mu\nu}(x^\mu) +\Delta \mathcal L(x^\mu)+\phi\left(\nabla_\mu\nabla_\nu T^{\mu\nu} + \frac{\Lambda}{3} T\right)\right],
\end{equation}
where $\Delta T_{\mu\nu}=T_{\mu\nu}(\Phi^\prime_I)-T_{\mu\nu}(\Phi_I)$ and $\Delta \mathcal L(x^\mu)=\mathcal L(\Phi^\prime_I)-\mathcal L(\Phi_I)$. 

Even though we start with local parameters of $\phi$ and $\zeta_I$, at least one of them becomes nonlocal to achieve the invariance. Namely,
\begin{equation}
\phi=-\frac{h_{\mu\nu}(x^\mu)\Delta T^{\mu\nu}(x^\mu) +\Delta \mathcal L(x^\mu)}{\nabla_\mu\nabla_\nu T^{\mu\nu} + \frac{\Lambda}{3} T},
\end{equation}
where on the denominator of the transform rule contains covariant derivatives. The inverse of derivatives 
can be interpreted as an integration, which means that it has a nonlocal term.

Maybe we cannot avoid the nonlocality for the construction of the symmetry transform\fixmeag{ation}{}. Therefore, we develop the nonlocal property of the transformation and see how the situation goes on.

\paragraph{The matter fields transformation construction} To be more precise, consider massless scalar fields coupled to PM gravity theory. Its kinetic part of the Lagrangian density and interaction with the the gravitational fields are given by
\begin{equation}
S_M[\Phi,h_{\mu\nu}]=\int \sqrt{-g}d^4x \partial_\mu \Phi\left(g^{\mu\nu}+ h^{\mu\nu}-\frac{1}{2}hg^{\mu\nu}\right) \partial_\nu \Phi \\ 
\end{equation}

Now we consider a QED like field transformation. Under the $U(1)$ gauge transformation, the phase of the fermionic fields changes in QED. However, the fields in this game are all real. Therefore, we demand that
\begin{equation}
\partial_\mu \Phi^\prime(x^\alpha)= M_\mu^{\ \nu}(x^\alpha)\partial_\nu \Phi(x^\alpha),
\end{equation}
where $M_\mu^{\ \nu}(x^\alpha)$ is arbitrary 4$\times$4 matrix which to be determined. Absolutely this causes nonlocal transform\fixmeag{ation}{} of the fundamental field $\phi$ as
\begin{equation}
\ \Phi^\prime(y^\alpha)= \int^{y^\alpha}M_\mu^{\ \nu}(x^\alpha)\partial_\nu  \Phi(x^\alpha)dx^\mu,
\end{equation}
or equivalently
\begin{equation}
\ \Phi^\prime(x^\alpha)=\nabla^{-2}\nabla^\mu(M_\mu^{\ \nu}(x^\alpha)\partial_\nu \Phi(x^\alpha))
\end{equation}
At this stage, just consider its infinitesimal transformation as 
\begin{equation}
M_\mu^{\ \nu}=\delta_\mu^{\ \nu}+\epsilon_\mu^{\ \nu}
\end{equation}
The condition of invariance of the matter action provides that
\begin{equation}
0=\int \sqrt{-g}d^4x \partial^\mu \Phi\partial^\nu \Phi \left(
\nabla_\mu\nabla_\nu \phi -\frac{1}{2} g_{\mu\nu}\left(\nabla^2\phi+\frac{2\Lambda}{3} \phi\right)+2\epsilon_\mu^{\ \alpha}\left(g_{\alpha \nu}+ h_{\alpha \nu}-\frac{1}{2}hg_{\alpha \nu}\right)
\right)
\end{equation}
The solution of this condition is given by
\begin{equation}
\epsilon_\mu^{\ \nu}=[K^{-1}]_\mu^{\ \alpha}\left[
\nabla_\alpha\nabla^\nu \phi -\frac{1}{2} g_{\alpha}^{\ \nu}\left(\nabla^2\phi+\frac{2\Lambda}{3} \phi\right)\right],
\end{equation}
where
\begin{equation}
K_{\alpha\nu}=2\left(g_{\alpha \nu}+ h_{\alpha \nu}-\frac{1}{2}hg_{\alpha \nu}\right)
\end{equation}

From the analysis above, we have seen, at least in the case of the scalar matter field, that we might be able to realize a QED-like transformation on the matter field according to the PM transformation. 

It is possible to extend this analysis to the case of a vector matter field. Consider the case where PM gravity is sourced by a vector field as the following,
\begin{eqnarray}
S= S_{PM}+\int \sqrt{-g}d^4x h_{\mu\nu}(x^\mu)T^{\mu\nu}(x^\mu) -\frac{1}{4}\mathcal F_{\mu\nu}\mathcal F^{\mu\nu}, \label{vectormatter}
\end{eqnarray}
where the energy-momentum tensor now takes the form
\begin{eqnarray}
T_{\mu\nu}=-\frac{1}{2}\left( \mathcal F_{\mu\alpha}\mathcal F_\nu^{\ \alpha} - \frac{1}{4} g_{\mu\nu}\mathcal F_{\alpha\beta}\mathcal F^{\alpha\beta} \right),
\end{eqnarray}
and $\mathcal F_{\mu\nu}\equiv\partial_\mu A_\nu-\partial_\nu A_\mu$.
Under the PM transformation, suppose that the vector field transforms according to the following:
\begin{eqnarray}
\partial_\mu A'_\nu = {M_{\mu\nu}}^{\alpha\beta} \partial_\alpha A_\beta.
\end{eqnarray}
Similar to the scalar field case, this transformation results in a nonlocal transformation on the field $A_\mu$. Now suppose that ${M_{\mu\nu}}^{\alpha\beta}$ can be expressed infinitesimally as
\begin{eqnarray}
{M_{\mu\nu}}^{\alpha\beta}=\delta^{[\alpha}_\mu\delta^{\beta]}_\nu + {\epsilon_{[\mu\nu]}}^{[\alpha\beta]}.
\end{eqnarray}
Under such a change, the field strength tensor transforms as
\begin{eqnarray}
\mathcal F'_{\mu\nu}(x^\alpha)=\left(\delta_\mu^{[\alpha}\delta_\nu^{\beta]}+{\epsilon_{[\mu\nu]}}^{[\alpha\beta]}\right)\mathcal F_{\alpha\beta}(x^\alpha),
\end{eqnarray}
where $A^{[\alpha}B^{\beta]}\equiv(A^{\alpha}B^{\beta}-A^{\beta}B^{\alpha})/2$.
We require that the interaction Lagrangian density, (\ref{vectormatter}) is invariant under this transform\fixmeag{ation}{} together with the PM transformation, which gives th\fixmeag{e}{at} following condition,
\begin{eqnarray}
0=\frac{1}{2}\int\sqrt{-g}d^4x\mathcal F_{\mu\nu}\left[-\delta^{[\nu}_{[\sigma} \nabla^{\mu]}\nabla_{\rho]} \phi+\frac{1}{4}\delta^{[\mu}_\rho\delta^{\nu]}_\sigma\nabla^2\phi-2{\epsilon_{[\alpha\sigma]}}^{[\mu\nu]}h^\alpha_\rho +{\epsilon_{[\rho\sigma]}}^{[\mu\nu]}\left(\frac{h}{2}-1\right)\right]\mathcal F^{\rho\sigma}.
\end{eqnarray}
\fixmeag{From the above equation, t}{T}he solution for ${\epsilon_{\rho\sigma}}^{\mu\nu}$ is 
\begin{eqnarray}
{\epsilon_{[\rho\sigma]}}^{[\mu\nu]}={[B^{-1}]_{[\rho}}^\alpha\left(\nabla^{[\mu}\nabla_\alpha \delta^{\nu]}_{\sigma]}-\frac{1}{4}\delta^{[\mu}_\alpha\delta^{\nu]}_{\sigma]}\nabla^2\right)\phi,
\end{eqnarray}
where
\begin{eqnarray}
B_{\mu\nu}\equiv\left(\frac{h}{2}-1\right)g_{\mu\nu}-2h_{\mu\nu}.
\end{eqnarray}

\paragraph{Drawbacks}
In this subsection, we show that the matter fields coupled to the PM-gravity fields in a way that $\mathcal L_{int}\sim h_{\mu\nu} T^{\mu\nu}$ necessarily transform non-locally. There are several pieces of criticism on the non-local transformation. First, somehow if the non-local transformation is allowed, then the gauge symmetry and the corresponding Noether process
are trivially achieved. Allowing Non-locality is not as much restrictive as local-transformation. The locality gives the genuine and strong constraints on the gauge theory and the theory becomes rather refined. Second, non-locality can cause non-unitarity and/or a-causality. These are typical obstructions for generalization of PM-gravity theory to higher order (self)-interactions. There are several examples as such. In \cite{Joung:2014aba}, The authors consider an interaction between the two fields in the theory: the partially massless spin 2 fields and the gravitational fields from Einstein-Hilbert theory. They find that the theory becomes non-unitary since it presents either the relative sign in the kinetic terms of the two fields are opposite or the couplings between the two fields become imaginary.
In \cite{Apolo:2016ort}, the authors construct a non linear conformal gravity based on $SO(1,5)$ group together with lcoal $U(1)$ symmetry. They introduce two different vielbeins, and those are complexifed. The $U(1)$ group mediates rotation of doublet of the two vielbeins. The theory obtained provides partially massless theory in the linearized level and address full order interactions for it. The partially massless $U(1)$ symmetry is originated from the extra $U(1)$ of the conformal gravity. However, they do not study the ghost issues for the full order of the transformations.
In \cite{Garcia-Saenz:2015mqi}, The authors prove that a partially massless Yang–Mills theory does not exist.

\subsection{Gauge covariant derivatives}
To overcome the drawbacks discussed in the previous chapter, we suggest an alternative way to construct gravity-matter fields interactions which is $\bf local$ transformation. We consider gauge covariant derivatives where the partially-massless gauge transformation is related to a phase change of the matter(complex) fields. For this, we need to develop an appropriate gauge connection which removes terms originated from the change of phase of the matter fields. Fortunately, we have such objects in PM-gravity theory.

Spin 1 objects being made out of $h_{\mu\nu}$ containing at most a single derivative are the following two:
\begin{equation}
\nabla_\nu h^{\mu\nu} {\ \ \rm and \ \ }\nabla^\mu h^\alpha_\alpha.
\end{equation}
Especially, among their linear combinations, the particular one as
\begin{equation}
\Gamma^{(0)}_\mu =\nabla^{\nu}h_{\mu\nu}-\nabla_{\mu}h
\end{equation}
vanishes in the level of equation of motion of (free)PM gravity theory\cite{Hinterbichler:2014xga}, which might ensure current conservation of matter fields coupled to PM gravity fields. Therefore, considering the following objects is to be reasonable;
\begin{equation}
\Gamma^{(1)}_\mu(\alpha,h_{\mu\nu})\equiv \nabla^{\nu}h_{\mu\nu}-\alpha\nabla_{\mu}h,
\end{equation}
where the $\alpha\neq 1$. Once we take into account the null quantity $\Gamma^{(0)}_\mu$, then the object $\Gamma^{(1)}_\mu(\alpha,h_{\mu\nu})$ effectively becomes $\Gamma^{(1)}_\mu(\alpha,h_{\mu\nu})=(1-\alpha)\nabla^{\nu}h_{\mu\nu}$. Then, here we examine $\alpha=0$ case, $\Gamma^{(1)}_\mu=\nabla^{\nu}h_{\mu\nu}$ only and it will be enough.

It turns out that under the partially-massless $U(1)$-transformation, it transforms as
\begin{equation}
\delta \Gamma^{(1)}_\mu=  \nabla_\mu \beta_1,
\end{equation}
where $\beta_1$ is a scalar quamtity being given by
\begin{equation}
\beta_1=\nabla^2\phi +\frac{4\Lambda}{3}\phi
\end{equation}
Therefore, this can be a candidate for the $U(1)$-gauge connection.
By using this fact, one can construct a gauge covariant derivative acting on complex matter fields. We define the covariant derivative as
\begin{equation}
D^{(1)}_\mu=\partial_\mu+i\Gamma^{(1)}_\mu.
\end{equation}
Suppose there is (complex)matter fields transforming as
\begin{equation}
\zeta \rightarrow \zeta e^{-i\beta_1}, \zeta^* \rightarrow \zeta^* e^{i\beta_1},
\end{equation}
then we see 
\begin{equation}
D^{(1)\prime}_\mu\zeta' \rightarrow e^{-i\beta_1} D^{(1)}_\mu\zeta.
\end{equation}

Second of all, there is another object being worthy to consider. 
Let us look at the following object.
\begin{equation}
\Gamma^{(2)}\equiv \nabla_\mu\nabla_\nu h^{\mu\nu},
\end{equation}
Under the partially-massless $U(1)$ gauge transformation, this object $\Gamma^{(2)}$ transforms as
\begin{equation}
\delta \Gamma^{(2)}=\nabla^2 \beta_1 \equiv \beta_2.
\end{equation}
In this case, an object,
\begin{equation}
\eta e^{i\kappa^2_4 \Gamma^{(2)}} 
\end{equation}
becomes invariant under the partially-massless $U(1)$ gauge transformation if the matter fields' transformation of $\eta$ is given by
\begin{equation}
\eta \rightarrow \eta e^{-i\kappa_4^2\beta_2}, \eta^* \rightarrow \eta^* e^{i\kappa_4^2\beta_2}
\end{equation}
where $\kappa_4$ is the 4-dimensional gravity constant, i.e. $\kappa_4=\sqrt{8\pi G}$. We introduce the $\kappa_4$ by dimensional analysis argument, which is the coupling constant of gravity theory.

Then, a factor $\partial_\mu( \eta e^{i\kappa^2_4 \Gamma^{(2)}} )$ can be written in terms of another type of gauge covariant derivative.
\begin{equation}
\partial_\mu\left( \eta e^{i\kappa^2_4 \Gamma^{(2)}} \right)\rightarrow e^{i\kappa^2_4 \Gamma^{(2)}}D^{(2)}_\mu \eta=
 e^{i\kappa^2_4 \Gamma^{(2)}}\left( \partial_\mu +i\kappa^2_4 \Gamma^{(2)}_\mu\right)\eta,
\end{equation}
where the $\Gamma^{(2)}_\mu$ is given by
\begin{equation}
\Gamma^{(2)}_\mu=\nabla_\mu \Gamma^{(2)}.
\end{equation}
{\bf In the following, we will address possible matter field actions.}

\paragraph{Complex scalar field} Consider a complex scalar field which changes as
\begin{equation}
\label{complex-scalar-pm}
\Phi \rightarrow \Phi e^{-i\beta_1}, \Phi^* \rightarrow \Phi^* e^{i\beta_1}
\end{equation}
under partially massless $U(1)$ gauge transformation. Then, the first-type of the covariant derivative, $D^{(1)}_\mu\Phi$ transform is given by
\begin{equation}
D^{(1)}_\mu\Phi \rightarrow e^{-i\beta_1} D^{(1)}_\mu\Phi , D^{(1)*}_\mu\Phi^* \rightarrow e^{i\beta_1} D^{(1)*}_\mu\Phi^*.
\end{equation}
With such a covariant derivative, the matter action of this complex scalar is given by
\begin{equation}
S_\Phi=\int d^4x \sqrt{g}\mathcal L_{\Phi},
\end{equation} 
where the matter Lagrangian density is
\begin{equation}
\label{ms-1type}
\mathcal L_{\Phi}= \frac{1}{2}g^{\mu\nu}\left(D^{(1)}_\mu\Phi\right)^* D^{(1)}_\nu\Phi -\frac{1}{2}m^2_{\Phi} \Phi^*\Phi,
\end{equation}
$m_\Phi$ is the mass of the scalar field, and $g_{\mu\nu}$ is the de sitter space metric.
The second type of the covariant derivative for the complex scalar which transforms as
\begin{equation}
\Phi \rightarrow \Phi e^{-i\beta_2}, \Phi^* \rightarrow \Phi^* e^{i\beta_2},
\end{equation}
under the partially massless $U(1)$ gauge transformation provides the following matter Lagranfian density:
\begin{eqnarray}
\label{ms-2type}
\mathcal L_{\Phi}&=&\frac{1}{2}g^{\mu\nu}\partial_\mu\left( \Phi e^{i\kappa^2_4 \Gamma^{(2)}}  \right)
\partial_\nu\left( \Phi e^{i\kappa^2_4 \Gamma^{(2)}}  \right)^*-\frac{1}{2}m^2_{\Phi}\Phi\Phi^* \\ \nonumber
&=& \frac{1}{2}g^{\mu\nu}\left(D^{(2)}_\mu\Phi\right)^* D^{(2)}_\nu\Phi -\frac{1}{2}m^2_{\Phi} \Phi^*\Phi.
\end{eqnarray}

We note that the interaction vertex of $\mathcal L_{int}\sim i\kappa_4^2\Phi^*\Gamma^{(2)\mu} \partial_\mu \Phi+c.c$ from (\ref{ms-2type}) is higher order in derivative and higher power in the coupling of $\kappa_4$ than the vertax $\mathcal L_{int}\sim i\Phi^*\Gamma^{(1)\mu} \partial_\mu \Phi+c.c$ from (\ref{ms-1type}).

\paragraph{Fermionic fields} The fermionic field matter action can be constructed in the same manners with the gauge connections of $\Gamma^{(1)}_\mu$ and/or $\Gamma^{(2)}_\mu$. They will be given by
\begin{equation}
\mathcal L_\psi=i\bar \psi\gamma^\mu D^{(I)}_\mu\psi -m_\psi\bar\psi\psi,
\end{equation}
where $I=$ 1 or 2, $\gamma^\mu$ are the gamma metrices defined in de sitter space being given by $\gamma^\mu=\gamma^a e_a^\mu$ and $e_a^\mu$ is the vielbein.

\section{Solutions of the equations of motion and their quantization}
\label{Solutions of the equations of motion and their quantization}
\subsection{Mode decomposition and the equations of motion}
To \fixmeag{realize}{get} the actions and the equations of motion of \fixmeag{both the tensor mode and the vector mode of }{}the partially massless gravit\fixmeag{y}{ational fields}, we decompose the fields $h_{\mu\nu}$ as
\begin{eqnarray}
h_{0}^{\ 0}&=&2\Phi,\label{3+1decom1}
\\
h_{0}^{\ i}&=&B^T_i+\partial_i B,\label{3+1decom2}
\\
h_{i}^{\ j}&=&h^{TT}_{ij} +\frac{1}{2}\left(\partial_i E^T_j+\partial_j E^T_i\right) +2\delta_{ij}\Psi +\left(\partial_i\partial_j-\frac{1}{3}\delta_{ij}\partial_k\partial_k\right)E,\label{3+1decom3}
\end{eqnarray}
where $\partial_i B^T_i=0,\partial_i h^{TT}_{ij}=0,\delta_{ij}h^{TT}_{ij}=0,\partial_i E^T_i=0$, where we use $\delta_{ij}$ to contract the spacial indices $i,j$ and so on.

\subsection{The traceless tensor modes}
\paragraph{The tensor mode action}The tensor action is given by
\begin{eqnarray}
S_{T}=\int d^4x \frac{3}{\Lambda t^2}\left(\frac{1}{4}\dot{h}^{TT}_{ij}\dot{h}^{TT}_{ij}-\frac{1}{4}\partial_ih^{TT}_{jk}\partial_i h^{TT}_{jk}\right)-\frac{9}{4\Lambda^2t^4}m^2h^{TT}_{ij}h^{TT}_{ij}.\label{tensoraction}
\end{eqnarray}
The equation of motion for the tensor mode is
\begin{eqnarray}
\ddot{h}^{TT}_{ij}-\frac{2}{t}\dot{h}^{TT}_{ij}-\partial^k\partial_k h^{TT}_{ij}+\frac{3}{\Lambda t^2}m^2 h^{TT}_{ij}=0.
\end{eqnarray}
\paragraph{Mode solutions of the equations of motion}
We use $H_{ij}$ for $h^{TT}_{ij}$ for convenience.
The solution of the traceless tensor modes is given by
\begin{equation}
\label{Hij-solution}
H_{ij}(t,\vec k)=\int d^3 k \sum_{\sigma=\pm2}(a_{ij}(\sigma,k)u(t,k)+a^\dagger_{ij}(\sigma,k)u^\star(t,k)),
\end{equation}
where $a_{ij}$ is an arbitrary momentum-$\vec k$-depend\fixmeag{ent}{ing} coeffic\fixmeag{i}{}ent and $u(t,k)$ is the mode solution which is given by
\begin{equation}
u(t,k)=\frac{1}{(2\pi)^{\frac{3}{2}}\sqrt{2|k|}}\left(\sqrt{\frac{\Lambda }{3}}t\right)e^{-i|k|t-ik_ix_i}.
\end{equation}
The second term in the solution (\ref{Hij-solution}) is the Hermitian conjugate of the first term which ensures reality of the solution.
We define an inner product among them, which is given by
\begin{eqnarray}
(u(k,t),u(k^\prime,t))&=&-i\int d^3x\sqrt{h}{\ } n^\mu 
[u(k,t) \overleftrightarrow \partial_\mu u^\star(k^\prime,t)], \\
\nonumber
&=&\delta^{(3)}(k-k^\prime)
\end{eqnarray}
where 
$h$ is the determinant of the spacelike hypersurface of the metric $h_{ij}$, which is given by
\begin{equation}
ds^2_{\rm hs}=\frac{3}{\Lambda t^2}\sum_{i=1,2,3}(d x_i^2)
\end{equation}
and the timelike normal vector is chosen as
\begin{equation}
n^\mu=\sqrt{\frac{\Lambda}{3}}t{\ }\delta_t^\mu.
\end{equation}
\paragraph{Quantization of the tensor modes}
For the quantization of the traceless transverse tensor mode of the solutions, it is requested that
\begin{equation}
[H_{ij}(x,t),P_{kl}(x^\prime,t)]=if_{ij,kl}(x)\delta^{(3)}(x-x^\prime), \label{tensorcommute}
\end{equation}
where the $P_{ij}$ is the canonical conjugate of the fields $H_{ij}$ and
the structure constant $f_{ij,kl}$ has properties as
\begin{eqnarray}
f_{ij,kl}(x)=f_{ji,kl}(x)=f_{ij,lk}(x)=f_{kl,ij}(x), {\ \ }\partial_i f_{ij,kl}(x)=0, {\ \ \rm \ and\ \ } \delta_{ij}f_{ij,kl}(x)=0,
\end{eqnarray}
to ensure that the quantum fields are traceless and transverse.
In fact, the structure constant is given by
\begin{equation}
f_{ij,kl}(k)=2\left( \frac{1}{2}\Pi_{ij}\Pi_{kl} +\delta_{i(k}\Pi_{l)j}+\delta_{j(k}\Pi_{l)i}  - (\delta_{ij}\Pi_{kl}+\delta_{kl}\Pi_{ij})+\delta_{ij}\delta_{kl}-\delta_{i(k}\delta_{l)j}\right),
\end{equation}
in momentum space, where $\Pi_{ij}$ is the projection operator which is given by
\begin{equation}
\Pi_{ij}(k)=\delta_{ij}-\frac{k_i k_j}{k^2}.
\end{equation}

We define the creation and annihilation operators as
\begin{equation}
H_{ij}(x,t)=\int d^3 k \left(  a_{ij}(k)u(t,k)+a_{ij}^\dagger(k)u^\star(t,k)\right),
\end{equation}
then those satisfy the following commutation relation:
\begin{equation}
[a_{ij}(t,k),a^\dagger_{kl}(t,k^\prime)]=2f_{ij,kl}(k)\delta^{(3)}(k-k^\prime).
\end{equation}

The Hamiltonian operator in terms of the creation and annihilation operators is given by
\begin{equation}
H=\int d^3 |k|
(a_{ij}(k)a^\dagger_{ij}(k)+a^\dagger_{ij}(k)a_{ij}(k)).
\end{equation}
\paragraph{The two point correlators}
The two point correlation function of the transverse traceless tensor fields, $H_{ij}(k,t)$ is given by
\begin{equation}
\langle H_{ij}(t,k) H_{kl}(t^\prime,k^\prime)\rangle=\frac{4\Lambda}{3}tt^\prime \int \frac{d^4k}{2(2\pi)^3}\frac{f_{ij.kl}(k)e^{-ik_\mu (x-x^\prime)^\mu}}{\omega^2-k^2+i\epsilon}
\end{equation}

\paragraph{The tensor mode propagator in the frequency space}
We apply a Fourier transform\fixmeag{ation}{} from a real time to a frequency space by using the mode solution that we obtained previously as
\begin{equation}
H_{ij}(t,k)=\int \frac{d \omega}{(2\pi)^{1/2}}\sqrt{\frac{\Lambda}{6}}te^{-i\omega t}H_{ij}(\omega,k)
\end{equation}
To see the details of the divergence pieces from the time integral, we take its principle value as 
\begin{equation}
\int^0_{-\infty} dt \rightarrow \lim_{\epsilon \rightarrow 0, T \rightarrow \infty}\int^{-\epsilon}_{-T} dt
\end{equation}
Then, the tensor mode action $S_T$ is given by
\begin{eqnarray}
\nonumber
S_T&=& \lim_{\epsilon \rightarrow 0, T \rightarrow \infty}\int^{-\epsilon}_{-T} dt \int \frac{d\omega d\omega^\prime d^3k}{16\pi}
\left[-\omega\omega^\prime -k^2  -\frac{1}{t^2}-\frac{i(\omega+\omega^\prime)}{t}\right]e^{-i(\omega+\omega^\prime)t}H_{ij}(\omega,k)H_{ij}(\omega^\prime,-k), \\ \nonumber
&=& \int \frac{d\omega d\omega^\prime d^3k}{16\pi}\left[-(\omega\omega^\prime +k^2 )\pi\delta(\omega+\omega^\prime) -i\lim_{T\rightarrow \infty}\left( \frac{1-\cos(\omega+\omega^\prime)T}{\omega+\omega^\prime} \right)(\omega\omega^\prime +k^2 )-\lim_{\epsilon \rightarrow 0}\frac{1}{\epsilon}\right]
\\ 
&\times& H_{ij}(\omega,k)H_{ij}(\omega^\prime,-k)
\end{eqnarray}
Inside the square bracket, the first term being proportional to $\delta$-function is the usual kernel for the free propagator. The second term is unusual and we show that it vanishes. 

To compute the term, we utilize  stationary phase approximation. In the $T\rightarrow \infty$ limit, the function, $1-cos(\omega+\omega^\prime)T$ oscillates very rapid\fixmeag{ly}{} in $\omega+\omega^\prime$ and we assume that the other functions multiplied on that changes very slow\fixmeag{ly}{} relative to it. 
For the detailed discussion, we introduce the other variables for the frequency integration as
\begin{equation}
x\equiv \frac{\omega+\omega^\prime}{\sqrt{2}} {\rm\ \ and\ \ }y\equiv \frac{\omega-\omega^\prime}{\sqrt{2}}
\end{equation}
Since the Jacobian of the transformation is 1, the frequency integration of the second term is given by
\begin{eqnarray}
\label{the second ST}
-i
\int d\omega d\omega^\prime  \lim_{T\rightarrow \infty}\left( \frac{1-\cos(\omega+\omega^\prime)T}{\omega+\omega^\prime} \right)(\omega\omega^\prime +k^2 )H_{ij}(\omega,k)H_{ij}(\omega^\prime,-k)\\ \nonumber
=-i\lim_{T\rightarrow \infty}\int dx dy \left( \frac{1-\cos(\sqrt{2}xT)}{\sqrt{2}x} \right)\left(\frac{y^2-x^2}{2} +k^2 \right)H_{ij}(\omega,k)H_{ij}(\omega^\prime,-k).
\end{eqnarray}
The kernel of the integrand is an odd function in $x$ and we find the extrema of the function. First, find the extrema of $ \left( \frac{1-\cos(\sqrt{2}xT)}{\sqrt{2}x} \right)\left(\frac{y^2-x^2}{2} +k^2 \right)$ in $x$. The condition for this is approximately given by
\fixmeag{$\sqrt{2}xT=n\pi$}{$\sqrt{2}xT=\frac{2n+1}{2}\pi$} by assuming that the other functions are slowly varying where $n \in \mathcal Z$. Think of the nearest extrema from $x=0$,\fixmeag{ $n=1 \rm\ \  and\  -1$}{$n=0 \rm\ \  and\  -1$},  which are located at \fixmeag{$x=\pm\frac{\pi}{\sqrt{2}T}$}{$x=\pm\frac{\pi}{2\sqrt{2}T}$}. By using the stationary phase approximation, (\ref{the second ST}) becomes
\begin{eqnarray}
\nonumber
&\sim&-i\lim_{T\rightarrow \infty}\int dy \frac{\pi}{\sqrt{2}T}\left( \frac{1}{\frac{\pi}{T}} \right)\left(\frac{y^2-\left(\frac{\pi}{\sqrt{2}T}\right)^2}{2} +k^2 \right)H_{ij}(\omega,k)H_{ij}(-\omega-\frac{\pi}{2T},-k)\\ \nonumber
&-&i\lim_{T\rightarrow \infty}\int dy \frac{\pi}{\sqrt{2}T}\left( \frac{-1}{\frac{\pi}{T}} \right)\left(\frac{y^2-\left(\frac{\pi}{\sqrt{2}T}\right)^2}{2} +k^2 \right)H_{ij}(\omega,k)H_{ij}(-\omega+\frac{\pi}{2T},-k) \\ \nonumber
&\rightarrow& 0,
\end{eqnarray}
The same argument is applied to the other extrema and it turns out that they give null effect in the action. 
\paragraph{Adding boundary terms as a regulator}
The third term in the action is divergent as $\epsilon \rightarrow 0$. Even though it is divergent, a physical interpretation comes to understand that term. As the de Sitter space expands, the term proportional to cosmological constant and gravit\fixmeag{on}{ational} mass grows. In fact, they are proportional to its proper volume $\int \sqrt{-g} H_{ij}(t,x)H_{ij}(t,x) dt d^3x \sim t^{-1}V_3$, where $V_3$ is the coordinate volume of the space. 

However, such interpretation is inconsistent with the result in \cite{Deser:2001xr}. In \cite{Deser:2001xr}, the partially massless spin 2 fields travel in the speed of light. In fact, the infinite mass term $\sim \frac{1}{\epsilon}$ can be eliminated by adding appropriate boundary terms at both $t=-\epsilon$ and $t=-T$.
This is given by
\begin{equation}
S_{bdy}=-\left.\int \frac{d\omega d\omega^\prime d^3 k}{16\pi}\left(\frac{e^{-i(\omega+\omega^\prime)t}}{t}\right)H_{ij}(\omega,k)H_{ij}(\omega^\prime,-k)\right|^{t=-\epsilon}_{t=-T}.
\end{equation}

\subsection{The transverse vector modes}
\paragraph{The transverse vector modes action and their equations of motion}
The transverse vector modes action is given by
\begin{eqnarray}
S_V &=& \int d^4x \frac{3}{\Lambda t^2}\left(\frac{1}{8}\partial_i \dot{E}^T_j \partial_i \dot{E}^{T}_j +\frac{1}{2}\partial_i B^T_j \partial_i B^{T}_j +\frac{1}{2}\partial_i\partial_i \dot{E}^{T}_jB^T_j\right) \nonumber
\\
&&\quad -\frac{9m^2}{4\Lambda^2 t^4}\left(-2B^T_i B^{T}_i +\frac{1}{2}\partial_i E^T_j \partial_i E^{T}_j\right),\label{vectoraction}
\end{eqnarray}
and their equations of motion are
\begin{eqnarray}
\label{v11-equation}
-\partial_i\partial_i B^T_j +\frac{1}{2}\partial_i\partial_i \dot{E}^T_j +\frac{3}{\Lambda t^2}m^2 B^T_j =0,
\end{eqnarray}
and
\begin{eqnarray}
\label{v22-equation}
\partial_i\partial_i \ddot{E}^T_j -\frac{2}{t}\partial_i\partial_i \dot{E}^T_j -2\partial_i\partial_i \dot{B}^T_j +\frac{4}{t}\partial_i\partial_i B_j +\frac{3}{\Lambda t^2}m^2\partial_i\partial_i E^T_j =0,
\end{eqnarray}
where the dot, ``$\cdot$'' denotes the time derivative. \fixmeag{$B^T_j$}{$B^{T,j}$} is not a dynamical variable and we remove this field by using its equation of motion(\ref{v11-equation}). The action is that only for the field, $E^T_j$.
We obtain that action in momentum space by using Fourier transform as
\begin{eqnarray}
\label{Vector-actioN}
S_V &=& \int dt {d^3k}\left(\frac{3k^2}{\Lambda}\right)\left(\frac{1}{8t^2\left(1+\frac{k^2t^2}{2}\right)}\dot{E}_i(t,k)\dot{E}_i(t,-k)-\frac{1}{4t^4}E_i(t,k)E_i(t,-k)\right).\label{vecaction1}
\end{eqnarray}
and the equation of motion is
\begin{equation}
0=\frac{d}{dt}\left( \frac{\dot E_i(k,t)}{4t^2\left(1+\frac{k^2t^2}{2}\right)}\right)
+\frac{E_i(k,t)}{2t^4}.
\end{equation}
\paragraph{Null propagation of the vector modes}
As addressed in \cite{Deser:2001xr}, the partially massless fields travel in the speed of light. To see this, we define
\begin{equation}
{E}_i\equiv \tau\sqrt{\tau^2+2}\,\mathcal{E}_i, {\ \ \rm and\ \ }\mathcal{T}\equiv\sqrt{2}\tau+\tan^{-1}\frac{\sqrt{2}}{\tau}{\ \ \rm where\ \ }\tau\equiv \frac{kt}{\sqrt{2}},
\end{equation}
then the vector modes action transforms into a new field and time variable frame as
\begin{eqnarray}
S_V &=& \int d\mathcal{T} \frac{d^3k}{\left(2\pi\right)^3}\frac{1}{k}\left(\frac{3\sqrt{2}k^6}{32\Lambda}\right)\left(\frac{d\mathcal{E}_i(\mathcal{T},k)}{d\mathcal{T}}\frac{d\mathcal{E}_i(\mathcal{T},-k)}{d\mathcal{T}}-\mathcal{E}_i(\mathcal{T},k)\mathcal{E}_i(\mathcal{T},-k)\right). \label{vecaction5}
\end{eqnarray}
It is definite that the vector modes propagate along the null cone in this frame. This is due to the scale invariance, a particular direction of the de Sitter symmetry group. However, if one quantize this partially massless system, this i\fixmeag{s}{n} no longer true. Since the traceless tensor and the vector mode should share the same vacuum to create and annihilate their quanta. Since the Hamiltonian depends on the choice of time, in the time $t$-coordinate(conformal time), the vector mode is not light-like particle. In other words, in the time frame $\mathcal T$, the tensor modes are not light-like.

\paragraph{More on the vector modes} More careful observation on the vector mode action (\ref{Vector-actioN}), one realize\fixmeag{s}{} that its Lagrangian density is the same form as the tensor`s \fixmeag{up to}{upto} a time dependent overall factor multiplied on it. The overall factor is $\frac{1}{2+k^2t^2}$. Tensor  mode action can be mapped to a theory in the effectively flat space(in fact, it is a half of the flat space since $-\infty<t<0$). One can perform another field redefinition for the vector modes and find that it is a theory in the effective (half) flat space with time dependent potential. The transformation is given by
\begin{equation}
E_i(t,k)=t f(t) \bar E_i(t,k),{\rm \ \ where\ \ }f(t)=\sqrt{2+k^2t^2}
\end{equation}
With such a field redefinition, we get
\begin{eqnarray}
\nonumber
\label{the normalized vector action}
S_V&=&\int^{-\epsilon}_{-T} dtd^3k \left(\frac{3k^2}{\Lambda}\right)\left[ \frac{1}{4}\dot{\bar E}_i(t,k) \dot{\bar E}_i(t,-k)-\frac{k^2}{4}{\bar E}_i(t,k) {\bar E}_i(t,-k)\right. +\left.k^2V(|k|t){\bar E}_i(t,k) {\bar E}_i(t,-k)\right] 
\\ 
&+& \int d^3k \left(\frac{3k^2}{\Lambda}\right)\left.\left( \frac{1}{4t}+\frac{k^2 t}{4(2+k^2t^2)} \right){\bar E}_i(t,k) {\bar E}_i(t,-k)\right|^{t=-\epsilon}_{t=-T},
\end{eqnarray}
where the time dependent potential, $V(|k|t)$ is given by
\begin{equation}
V(|k|t)=\frac{2k^2t^2+1}{2(2+k^2t^2)^2}.
\end{equation}
The potential vanishes as $t \rightarrow -\infty$ and $V(0)=\frac{1}{8}$. By observing the Lagrangian density of the field $\bar E_{i}({t,k})$, it is massless field in the very early time(at $t=-\infty$). However, its dispersion relation may change as time goes by. In fact, in the very late time of the universe (near $t=0$), its dispersion relation becomes
\begin{equation}
\omega^2 = \frac{k^2}{{4}}.
\end{equation}
In fact, the solution of the equation of motion of the field $\bar E_i(t,k)$ can be obtained by looking at the solution of the primitive field, $E_i(t,k)$, which is
\begin{equation}
u_i(t,k)=\frac{2+i|k|t}{\sqrt{2+k^2t^2}}e^{-i|k|t} {\rm\ \ or \ \ its\  complex\  conjugate.}
\end{equation}
Since the factor in front of the exponential part in the solution is complex, it give\fixmeag{s}{n} extra phase. Therefore, we rewrite this in terms of its argument and modulus. Then, the solution becomes
\begin{equation}
u_i(t,k)=\frac{\sqrt{4+k^2t^2}}{\sqrt{2+k^2t^2}}\exp\left({-i|k|t+i\tan^{-1}\frac{|k|t}{2}} \right){\rm\ \ or \ \ its\  complex\  conjugate.}
\end{equation} 
Its instant frequency at the given time $t$ is obtained by $\omega(t)=\frac{d\Omega(t)}{dt}$, where the $\Omega=|k|t-\tan^{-1} \frac{|k|t}{2}$ is the phase factor of the solution. The instant frequency is given by
\begin{equation}
\omega(t)=|k|\frac{k^2t^2+2}{k^2t^2+4},
\end{equation}
which certify the dispersion relations in the very early and late time that we mention.

In the previous subsection, we discuss that the tensor mode's dispersion relation is that of a massless particle. This is probably right. The gravitational wave is observed recently and it travels (almost) at the speed of light. Partially massless gravity ensures this.

However, the vector modes show a different behavior. Its speed depends on the age of our Universe. If one can measure the speed of the vector parts of the gravitational waves, then one might know whether PM gravity is correct or ruled out. In other way to interpret it is that maybe PM gravity theory is correct and the dispersion relation tells us when we measure the vector particles. 

Another aspect of the vector mode is that the speed of the vector particle, \fixmeag{given by $v=\frac{d\omega}{d|k|}$}{}, depends on its momenta. We call this effect ``conformal prism" or ``de Sitter prism".
In fact, high frequency mode travels faster than the low frequency mode in a given time. 

Finally, the speed of the vector mode is bounded from below with $v=\frac{c}{{2}}$(50$\%$ of $c$), namely no less than that, where $c$ is the speed of light.

\paragraph{Mode solutions and their normalization}
The mode solution of the transverse vector modes are given by \fixmeag{
\begin{eqnarray}
v(t,k) = \frac{\sqrt{\Lambda}}{(2\pi)^{3/2}\sqrt{2|k|}} t \left(|k|t-2i\right)e^{-i|k|t-ik_ix_i}.
\end{eqnarray}
}{}
The mode solutions satisfy the normalization condition as
\begin{eqnarray}
(v(t,k),v(t,k^\prime))&=&-i\int d^3x \sqrt{h}n^\mu
[v(t,k)\overleftrightarrow\partial_\mu v^\star(t,k^\prime)]\rho(k,t)\\ \nonumber
&=&\delta^{(3)}(k-k^\prime)
\end{eqnarray}
where the nontrivial weight $\rho(k,t)=\frac{1}{1+\frac{k^2t^2}{2}}$. The most general solution of the vector fields are
\begin{eqnarray}
E_i(t,x) &=& \sum_{\sigma=\pm 1}\int d^3k \, a_i(k,\sigma) v(t,k) +  a_i^\dagger(k,\sigma)v^*(t,k),
\\
&\equiv& \int \frac{d^{3}k}{\left(2\pi\right)^{3/2}} \,e^{-ik_ix_i}E_i(t,k).
\end{eqnarray}
Namely, \fixmeag{
\begin{equation}
E_i(t,k)=\left(2\pi\right)^{3/2}\sum_{\sigma=\pm 1}[a_i(k,\sigma) v(t,k) +  a_i^\dagger(k,\sigma)v^*(t,k)]e^{ik_ix_i},
\end{equation}
}{}
where $E_i(t,k)$ is the solution in momentum space.

\paragraph{Qu\fixmeag{a}{}ntization of the vector modes}
The conjugate momentum for $E_i$ obtained by variation of the action (\ref{vecaction1}) is
\begin{eqnarray}
P_i(t,k) = \frac{2}{8t^2\left(1+\frac{k^2t^2}{2}\right)}\left(\frac{3k^2}{\Lambda}\right)\dot{E}_i(t,k).
\end{eqnarray}
We request the quantization rule as
\begin{equation}
\left[E_i(t,x),P_j(t,x')\right]=i\Pi_{ij}(x)\delta^{(3)}(x-x^\prime),
\end{equation}
which give rise to the commutation relations between the creation and the annihilation operators as
\begin{equation}
\left[a_i(k,\sigma),a_j^\dagger(k',\sigma')\right]=\frac{1}{4k^2}\delta_{\sigma\sigma^\prime} {\Pi}_{i\fixmeag{j}{}}(k)\delta^{(3)}(k-k')
\end{equation}
and the other commutators vanish.

One can define a creation and an annihilation operator spin summed as
\begin{equation}
a_i(k,t)=\sum_{\sigma=\pm 1}a_i(k,\sigma){\rm\ \ and\ \ }a^\dagger_i(k,t)=\sum_{\sigma=\pm 1}a^\dagger_i(k,\sigma),
\end{equation}
then they satisfy
\begin{equation}
\left[a_i(k,t),a^\dagger_j(k^\prime,t)\right]=\frac{1}{2k^2} {\Pi}_{ik}(k)\delta^{(3)}(k-k').
\end{equation}
\paragraph{Vector modes Hamiltonian and creation of the quanta}
The Hamiltonian obtained from the action(\ref{the normalized vector action}) by removing its boundary terms is given by
\begin{eqnarray}
H&=&\int d^3k 
\frac{3 |k|^3}{4}\left[\frac{\left(k^6t^6+4k^4t^4+5k^2t^2+6\right)}{(k^2t^2+2)^3} \left(a_i(k)a_i^\dagger(k)+a_i^\dagger(k)a_i(k)\right)\right.\nonumber
\\
&&+\frac{(5|k|^2t^2-2)+2i|k|t(k^2t^2-2)}{(k^2t^2+2)^3}e^{-2i|k|t}a_i(k)a_i(k)\nonumber
\\
&&\left.+\frac{(5|k|^2t^2-2)-2i|k|t(k^2t^2-2)}{(k^2t^2+2)^3}e^{2i|k|t}a_i^\dagger(k)a_i^\dagger(k)\right],
\end{eqnarray}
where we promote the $a(k)$ and $a^\dagger(k)$ to the annihilation and the creation operator respectively. The Hamiltonian is time dependent and in the very early time as $t=-\infty$, it is given by
\begin{equation}
H=\int d^3k 
\frac{3 |k|^3}{4} \left(a_i(k)a_i^\dagger(k)+a_i^\dagger(k)a_i(k)\right).
\end{equation}
However, in gneral, this becomes
\begin{equation}
H=\int d^3k 
\frac{3 |k|^3}{4} \left(b_i(k,t)b_i^\dagger(k,t)+b_i^\dagger(k,t)b_i(k,t)\right),
\end{equation}
where
\begin{eqnarray}
b(t,k)=F(t,k)e^{i\theta_F(t,k)}a(k)+G(t,k)e^{-i\theta_G(t,k)}a^\dagger(k), \\
b^\dagger(t,k)=F(t,k)e^{-i\theta_F(t,k)}a^\dagger(k)+G(t,k)e^{i\theta_G(t,k)}a(k)
\end{eqnarray}
The positive real functions $F(t,k),G(t,k)$ and the real functions $\theta_F(t,k), \theta_G(t,k)$ are given by
\begin{eqnarray}
\theta_F(t,k)+\theta_G(t,k)=-2|k|t+\tan^{-1}\left(\frac{2|k|t(k^2t^2-2)}{5k^2t^2-2}\right), \\
F^2(t,k)+G^2(t,k)=\frac{\left(k^6t^6+4k^4t^4+5k^2t^2+6\right)}{(k^2t^2+2)^3}, \\
F^2(t,k)G^2(t,k)=\frac{4k^6t^6+9k^4t^4-4k^2t^2+4}{4(k^2t^2+2)^6}.
\end{eqnarray}
Finally, the commutator between $b(k,t)$ and $b^\dagger(k,t)$ are given by
\begin{equation}
\left[b_i(k,t),b^\dagger_j(k^\prime,t)\right]=\frac{F^2(t,k)-G^2(t,k)}{2k^2} {\Pi}_{ik}(k)\delta^{(3)}(k-k').
\end{equation}

\section{Conclusions and Outlooks}
\subsection{Conclusion}
In this note we study the partially-massless gravity in 4 dimensions in the consideration of interactions with the quantum matter field. It is widely argued \cite{deRham:2013wv} that there is the ``no-go'' result showing that it is impossible to maintain the $U(1)$ symmetry in (\ref{PMsym}), a signature to the PM gravity, when one tries to add the PM theory the higher order gravitational self interaction. Alternatively, we suggest adding interactions with  matter field. The matter itself needs to satisfy the conservation law (\ref{current-con}) otherwise the theory cannot maintain the $U(1)$ PM symmetry. In other words, (\ref{current-con}) serves as an on-shell condition for the matter field. If one wants to relax this extra condition in order to construct a quantum theory out of the PM gravity, one needs to allow the matter to transform accordingly. We found that the matter transformation must be nonlocal. As we have tried with scalar matter field and vector matter field, we found that the infinitesimal transformations for both of the matters involve nonlinear contributions in the PM field $h_{\mu\nu}$. We also tried to find a general structure of a scalar matter which transforms accordingly under the PM transformation.

To realize its quantum theory of the PM gravity, we consider first its structure through equations of motion. By separating the PM field, $h_{\mu\nu}$, into distinctive modes, we evaluated mode solutions for both transverse-traceless tensor mode and transverse vector mode while the scalar mode of $h_{\mu\nu}$ decouples thanks to the PM symmetry. For the transverse-traceless tensor mode, we request the commutation relation (\ref{tensorcommute}) in order to quantize the tensor mode with its mode solution. We can define its creation and annihilation operators along with their commutation relation. Moreover, the tensor mode can be realized to have a null propagation in the frame where we use the conformal time $t$ (in de Sitter space). For the transverse vector mode, the situation is rather different. One distinct nature of the vector mode is that it propagates with speed of light in a different frame compared to the tensor case; in the frame described by the time coordinate $\mathcal{T}$ the vector mode has null propagation while it develops a nontrivial potential in the frame corresponding to the time coordinate $t$ where tensor mode has null propagation. These results explain that both modes can have null propagation in each of their appropriate frames but when one wants to quantize both modes on the same vacuum state simultaniously, one will face a situation where only one of them has null propagation. Furthermore, the vector mode seems to have a time-dependent dispersion relation which results in having null propagation at the very early time and slower propagation at late time. Interestingly, the dispersion relation of the vector mode also has momentum dependence, resulting in the speed of the vector particle depending on its momentum, dubbed in this study as ``comformal prism''. On the quantization aspect, by following the same procedures as done in the tensor mode case, we also are able to realize its creation and annihilation operators along with the corresponding commutation relation. By constructing the vector Hamiltonian, it is obvious that the Hamiltonian has nontrivial time dependences while with appropriate redefinitions the Hamiltonian can also be written in a standard form, suggesting that the vector mode can have null propagation in a system with a vacuum state different from that of the tensor mode. 

\subsection{Outlooks}
\paragraph{More studies on the matter fields} 
The essential spirit of the theory of general relativity is the ``equivalence principle." The reason why the gravity can be described by geometry is due to this property. The quantum mechanical realization of the equivalence principle is given by introducing interactions to the matter fields in such a way that the gravity field couples to the stress energy tensor of the matter fields with the same interaction strength regardless of their kinds.

In this note, we consider the partially massless gravity theory as a serious candidate of quantum gravity theory.  
To discuss the equivalence principle in this context, we must consider an appropriate coupling to the matter field. The interaction is given by $\sim \kappa_4 \sqrt{g}h_{\mu\nu}T^{\mu\nu}$ as explained, where $\kappa_4$ is the 4-dimensional gravity constant.
The matter field would also be a quantum matter since we want to study a full quantum theory. 

We do not request any constraints to the stress energy tensor as stress energy tensor conservation (\ref{extra-u1-condition}) since it is an on-shell condition. Rather than this, we realize that under the U(1) transformation of the partially massless gravity theory, the matter field must transform accordingly. In fact, we develop the infinitesimal transformation of the matter fields especially for the massless scalar and vector gauge fields in this paper. 

We will ask more questions on this issue as
\begin{itemize}
\item What is the finite version of the matter field transformation?
\item What is the Noether charge and the conservation current of the symmetry?
\item What is the corresponding Ward identity(or Schwinger-Dyson equation)? 
\item Any anomalies?
\end{itemize}

\paragraph{No real life gravity without matters?}
Considering the matter fields as quantum fields implies that we keep internal lines or(and) internal loops of the matter fields in mind when we compute the correlation functions of the gravitational fields. Somehow this open to possibilities such that the the gravitational effects that we observe do not just come from the gravitational fields themselves but they are also affected by an internal mediation by the matter fields. In our model, we consider couplings between gravitational fields and the matter fields and those kinds only. Therefore, we expect that we can see such matter fields contributions to our quantum calculations. 

\paragraph{Einstein gravity as an effective action}
We will try to perform quantum calculations of the partially massless gravity theory to present their (at least 1-loop) effective action. Einstein's theory of general relativity is well designed classical theory and this is probably comprised of a kind of collective effects of a certain quantum gravity theory. It might be good if we compute every possible quantum effective interactions and realize that the collection of such interaction give rise to Einstein gravity or the similar.


\section*{Acknowledgement}
J.H.O Thanks Yoonbai Kim, Mu-in Park, Yunsoo Myung and Dong-Han Yeom for useful discussion in ``STGCOS 2018''(String theory, Gravitation and Cosmology 2018) in Pohang. J.H.O especially thank his W.J. and Y.J. This work was supported by the National Research Foundation of Korea(NRF) grant funded by the Korea government(MSIP) (No.2016R1C1B1010107) and Research Institute for Natural Sciences, Hanyang University.

\appendix
\section{Appendices}
\subsection{Decomposition of the modes of gravitational fields in generic D dimensions}
The theory that we consider is \fixme{the}{that} massive Fierz-Pauli action coupled  to \fixme{a}{} $U(1)$-gauge field. The action is given by
\begin{eqnarray}
\nonumber
S&=&\int d^D x \sqrt{-g}\left[  -\frac{1}{2}\nabla_\lambda h_{\mu\nu} \nabla^\lambda h^{\mu\nu} + \nabla_\lambda h_{\mu\nu} \nabla^\nu h^{\mu\lambda}
-\nabla_\mu h \nabla_\nu h^{\mu\nu} +\frac{1}{2}\nabla_\mu h \nabla^\mu h   \right. \\ 
&&\quad-\frac{R}{2}\left( h^{\mu\nu}h_{\mu\nu}-\frac{1}{2}h^2 \right)+2R_{\mu\nu}\left(h^{\mu\rho}{h_\rho}^\nu-\frac{1}{2}hh^{\mu\nu}\right)+\Lambda\left( h^{\mu\nu}h_{\mu\nu}-\frac{1}{2}h^2 \right)
\\
&&\quad-\left.\frac{1}{2}m^2(h^{\mu\nu}h_{\mu\nu}-h^2)-\frac{1}{4}\mathcal F_{\mu\nu}\mathcal F^{\mu\nu}+h_{\mu\nu}T^{\mu\nu}\right],\nonumber
\end{eqnarray}
where $h_{\mu\nu}$ is a symmetric (real) tensor and we discuss this in the $D$-dimensional spacetime where $h_{\mu\nu}$ itself has $\frac{D(D+1)}{2}$ degrees of freedom(dof) (its dof in 4-dimension is 10). To analyze this theory, we employ Stuckelburg trick as
\begin{equation}
h_{\mu\nu} \rightarrow h_{\mu\nu}+ \nabla_\mu A_\nu+ \nabla_\nu A_\mu + 2\nabla_\mu\nabla_\nu \phi + \frac{2m^2}{D-2}\phi g_{\mu\nu}, \label{rep}
\end{equation}
where the (real)vector fields $A_\mu$ and the (real)scalar field $\phi$ are called the Stuckelburg fields. The  $A_\mu$ and $\phi$ are the most general fields(no restrictions on them yet). Therefore,  $A_\mu$ contain $D$ dofs and the $\phi$ has 1 dof. Moreover, the last part in Eq. (\ref{rep}) is introduced to diagonalize the $h-\phi$ kinetic mixing in the action and to reveal the $\phi$ kinetic term. 
The \fixme{right side of Eq. (\ref{rep})}{first line in the action} is invariant under the following transform:
\begin{eqnarray}
\delta h_{\mu\nu}&=&2\nabla_{(\mu} \xi_{\nu)}+\frac{2m^2}{D-2} \lambda g_{\mu\nu} \\ 
\delta A_\mu&=&-\xi_\mu+\nabla_\mu \lambda \\
\delta \phi&=&-\lambda
\end{eqnarray}
\fixme{Moreover, the $A_\mu$-replacement in Eq. (\ref{rep}) leaves the first two lines of the action invariant (on-shell).}{} 
Therefore, once we replace $h_{\mu\nu}$ by the Stuckelburg fields, the replaced parts become $h_{\mu\nu}+\frac{2m^2}{D-2}\phi \bar g_{\mu\nu}$.

According to the Stuckelburg trick, we compute each term in the first line in the action as follows,
\begin{eqnarray}
\nabla_\lambda h_{\mu\nu} \nabla^\lambda h^{\mu\nu} &\rightarrow& \nabla_\lambda (h_{\mu\nu}+ \frac{2m^2}{D-2}\phi  g_{\mu\nu}) \nabla^\lambda (h^{\mu\nu}+\frac{2m^2}{D-2}\phi  g^{\mu\nu})\\ \nonumber
&=&\nabla_\lambda h_{\mu\nu} \nabla^\lambda h^{\mu\nu}+ D\frac{4m^4}{(D-2)^2}\nabla_\lambda \phi \nabla^\lambda \phi +2\frac{2m^2}{D-2}\nabla_\lambda \phi \nabla^\lambda h, \\
\nabla_\lambda h_{\mu\nu} \nabla^\nu h^{\mu\lambda} &\rightarrow& \nabla_\lambda (h_{\mu\nu}+ \frac{2m^2}{D-2}\phi  g_{\mu\nu}) \nabla^\nu (h^{\mu\lambda}+\frac{2m^2}{D-2}\phi  g^{\mu\lambda})\\ \nonumber
&=&\nabla_\lambda h_{\mu\nu} \nabla^\nu h^{\mu\lambda}+ \frac{4m^4}{(D-2)^2}\nabla_\lambda \phi \nabla^\lambda \phi +2\frac{2m^2}{D-2}\nabla_\lambda \phi \nabla_\mu h^{\lambda\mu}, \\
\nabla_\mu h \nabla_\nu h^{\mu\nu} &\rightarrow& \nabla_\mu (h+D\frac{2m^2}{D-2}\phi) \nabla_\nu (h^{\mu\nu}+\frac{2m^2}{D-2}\phi g^{\mu\nu}) \\ \nonumber
&=&\nabla_\mu h \nabla_\nu h^{\mu\nu}+D\frac{4m^4}{(D-2)^2}\nabla_\lambda \phi \nabla^\lambda \phi+\frac{2m^2}{D-2}\nabla_\lambda h \nabla^\lambda \phi+D\frac{2m^2}{D-2} \nabla_\lambda \phi \nabla_\mu h^{\lambda\mu},\\ 
\nabla_\mu h \nabla^\mu h &\rightarrow& \nabla_\mu (h+D\frac{2m^2}{D-2}\phi) \nabla^\mu (h+D\frac{2m^2}{D-2}\phi) \\ \nonumber
&=& \nabla_\mu h \nabla^\mu h+ D^2\frac{4m^4}{(D-2)^2}\nabla_\mu \phi \nabla^\mu \phi+2D\frac{2m^2}{D-2}\nabla_\mu \phi \nabla^\mu h,
\end{eqnarray}
and
\begin{eqnarray}
h^{\mu\nu}h_{\mu\nu} &\rightarrow& (h^{\mu\nu}+\frac{2m^2}{D-2}\phi g^{\mu\nu})(h_{\mu\nu}+\frac{2m^2}{D-2}\phi g_{\mu\nu}) \\ \nonumber
&=&h^{\mu\nu}h_{\mu\nu}+D\frac{4m^4}{(D-2)^2}\phi^2+2\frac{2m^2}{D-2}\phi h \\
h^2 &\rightarrow& (h+D\frac{2m^2}{D-2}\phi)^2 = h^2+D^2\frac{4m^4}{(D-2)^2}\phi^2+2D\frac{2m^2}{D-2}\phi h.
\end{eqnarray}
By utilizing the above facts, the first \fixme{two lines}{line} in the action are given by
\begin{eqnarray}
\nonumber
S^\prime_{\rm massive-FP}&=&S_{\rm massive-FP}(h_{\mu\nu})
\\
&&+\int d^Dx \sqrt{-g} \left[  \frac{4m^4}{(D-2)^2} \left( \frac{D^2}{2}-\frac{3D}{2}+1 \right) \nabla_\lambda  \phi \nabla^\lambda \phi \right.\nonumber
\\
&&+\frac{4m^4}{(D-2)^2}\left(1-\frac{D}{2}  \right)\left(\left(2-\frac{D}{2}\right)R+\Lambda D\right)\phi^2 \nonumber
\\
&&-2m^2\left(  \nabla_\mu  \phi \nabla_\nu h^{\mu\nu}- \nabla_\lambda  \phi \nabla^\lambda h\right) - \left(4-D\right)\frac{m^2}{D-2}Rh\phi \nonumber
\\
&&+\left.\frac{2m^2}{D-2}(4-D)R_{\mu\nu}h^{\mu\nu}\phi -2m^2\Lambda h\phi\right]{ \ \ \ \phi h- \rm\bf \ mixing\ terms}
\end{eqnarray}
The second line of the action is expanded as belows.
The last term in the action is given by 
\begin{eqnarray}
h_{\mu\nu}T^{\mu\nu} &\rightarrow& \left(h_{\mu\nu}+ \nabla_\mu A_\nu+ \nabla_\nu A_\mu + 2\nabla_\mu\nabla_\nu \phi + \frac{2m^2}{D-2}\phi g_{\mu\nu}\right)T^{\mu\nu},
\\ \nonumber
&=&h_{\mu\nu}T^{\mu\nu} + 2\nabla_\mu A_\nu T^{\mu\nu} + 2\nabla_\mu\nabla_\nu \phi T^{\mu\nu} +\frac{2m^2}{D-2}\phi T,
\end{eqnarray}
where $T\equiv T^\mu_\mu $. The mass term is given by
\begin{eqnarray}
(h_{\mu\nu}h^{\mu\nu}-h^2) &\rightarrow&(h_{\mu\nu}h^{\mu\nu}-h^2) +\{(\nabla_\mu A_\nu+ \nabla_\nu A_\mu)^2-4(\nabla_\mu A^\mu)^2\} \\ \nonumber
&&+\{(2\nabla_\mu\nabla_\nu \phi+\frac{2m^2}{D-2}\phi g_{\mu\nu})^2-(2\nabla^2\phi + D\frac{2m^2}{D-2}\phi)^2\} \\ \nonumber
&&+2h^{\mu\nu}(\nabla_\mu A_\nu+ \nabla_\nu A_\mu)-4h\nabla_\mu A^\mu  { \ \ \ hA- \rm\bf \ mixing\ terms} \\ \nonumber
&&+ 4(2\nabla_\mu A_\nu \nabla^\mu \nabla^\nu \phi+\frac{2m^2}{D-2} \phi \nabla_\mu A^\mu) \\ \nonumber
&&-4(2\nabla_\mu A^\mu \nabla^2 \phi+D\frac{2m^2}{D-2} \phi \nabla_\mu A^\mu){ \ \ \ \phi A- \rm \ \bf mixing\ terms} \\ \nonumber
&&+2\left(2h^{\mu\nu}\nabla_\mu\nabla_\nu \phi+\frac{2m^2}{D-2}\phi h\right)\\ \nonumber
&&-2h(2\nabla^2\phi + D \frac{2m^2}{D-2}\phi){\;\;\qquad \qquad\ \ \ \phi h- \rm\bf \ mixing\ terms}
\end{eqnarray}
For the de Sitter space,
\begin{eqnarray}
R_{\mu\nu\rho\sigma}&=&\frac{2\Lambda}{(D-2)(D-1)}\left(g_{\mu\rho}g_{\nu\sigma}-g_{\mu\sigma}g_{\nu\rho}\right),
\\
R^\nu_{\ \beta\nu\mu}=R_{\beta\mu}&=&\frac{R}{D}g_{\beta\mu}=\frac{2\Lambda}{D-2}g_{\beta\mu},
\end{eqnarray}
where
\begin{equation}
\Lambda=\frac{D-2}{2D}R.
\end{equation}
By utilizing the fact that
\begin{equation}
\left(\nabla_\mu \nabla_\nu-\nabla_\nu \nabla_\mu\right)A^{\alpha}=R^{\alpha}_{\ \beta\mu\nu}A^\beta,
\end{equation}
one can show
\begin{equation}
\int d^Dx \sqrt{-g}\,\nabla_\mu A_\nu \nabla^\mu \nabla^\nu \phi = \int d^Dx \sqrt{-g}\left(\nabla_\mu A^\mu \nabla_\nu \nabla^\nu \phi-R^\nu_{\ \beta\nu\mu}A^\beta \nabla^\mu\phi\right) + {\rm total\ derivative\ terms},
\end{equation}
Then the { $\phi A-$\rm \ \bf mixing\ terms} are given by
\begin{eqnarray}
{ \ \ \phi A- \rm \ \bf mixing\ terms}&=&\int d^Dx\sqrt{-g}\left(-\frac{1}{2}m^2\right)4\left(2\nabla^\mu(A^\nu R_{\mu\nu})\phi+(1-D)\frac{2m^2}{D-2}\phi\nabla_\mu A^\mu\right)\nonumber
\\
&=&\int d^Dx\sqrt{-g}\left(-2m^2\right)\nabla_\mu A^\mu \phi\frac{2}{D-2}\left(2\Lambda - (D-1)m^2\right),\label{phiAmix}
\end{eqnarray}
Next we check the ${\  \phi h- \rm\bf \ mixing\ terms}$, which are given by
\begin{eqnarray}
{ \ \ \ \phi h- \rm\bf \ mixing\ terms}&=&\int d^Dx\sqrt{-g} \frac{2m^2}{D-2}h\phi\left(-2\Lambda+m^2\left(D-1\right)\right),\label{phihmix}
\end{eqnarray}
The $A_\mu$ kinetic and mass terms are
\begin{eqnarray}
{A_\mu-\rm kinetic\ terms} &\rightarrow& -\frac{m^2}{2}\int d^D x\sqrt{-g} \{(\nabla_\mu A_\nu+ \nabla_\nu A_\mu)^2-4(\nabla_\mu A^\mu)^2\} \\ \nonumber
&=&m^2\int d^D x\sqrt{-g}\left( -\frac{1}{2} F_{\mu\nu}F^{\mu\nu}+2\Lambda A_\mu A^\mu\right),
\end{eqnarray}
where for the second line, we use an identity as
\begin{equation}
\sqrt{-g}\nabla_\mu A^\mu\nabla_\nu A^\nu=\sqrt{-g}\nabla_\nu A^\mu\nabla_\mu A_\nu+
\sqrt{-g} R_{\mu\alpha}A^\mu A^\alpha + {\rm total\ derivatives},
\end{equation}
and $F_{\mu\nu}\equiv \nabla_\mu A_\nu- \nabla_\nu A_\mu $.

The scalar kinetic ter\fixme{m}{n}s are given by
\begin{eqnarray}
{\phi-\rm kinetic\ terms} &\rightarrow& \int d^Dx \sqrt{-g}\left(-m^2\left(D-1\right)+2\Lambda\right)\left( \frac{2m^2}{D-2}\nabla_\mu\phi\nabla^\mu\phi-\frac{D}{2} \frac{4m^4}{(D-2)^2}\phi^2 \right)\nonumber
\\ 
&+&{\rm total\ derivative\ terms}\label{phikin}
\end{eqnarray}
where we use the following identity:
\begin{equation}
\sqrt{-g} \nabla_\mu  \nabla_\nu \phi  \nabla^\mu  \nabla^\nu \phi = \sqrt{-g} \{(\nabla^2 \phi)^2 - R_{\mu\nu}\nabla^\mu\phi\nabla^\nu\phi \}+  {\rm total\ derivative\ terms}.
\end{equation}
To sum up, the whole action after the replacement is as follows,
\begin{eqnarray}
\nonumber
S&=&\int d^D x \sqrt{-g}\left[  -\frac{1}{2}\nabla_\lambda h_{\mu\nu} \nabla^\lambda h^{\mu\nu} + \nabla_\lambda h_{\mu\nu} \nabla^\nu h^{\mu\lambda}
-\nabla_\mu h \nabla_\nu h^{\mu\nu} +\frac{1}{2}\nabla_\mu h \nabla^\mu h   \right. \\ 
&-&\frac{\Lambda D}{D-2}\left( h^{\mu\nu}h_{\mu\nu}-\frac{1}{2}h^2 \right)+\frac{4\Lambda}{D-2}\left(h^{\mu\nu}{h_{\mu\nu}}-\frac{1}{2}h^2\right)+\Lambda\left( h^{\mu\nu}h_{\mu\nu}-\frac{1}{2}h^2 \right)\nonumber
\\
&-&\frac{1}{2}m^2(h^{\mu\nu}h_{\mu\nu}-h^2)-\frac{1}{4}\mathcal F_{\mu\nu}\mathcal F^{\mu\nu}+h_{\mu\nu}T^{\mu\nu} +2\nabla_{(\mu}A_{\nu)}T^{\mu\nu}+2\nabla_\mu\nabla_\nu\phi T^{\mu\nu}+\frac{2m^2}{D-2}\phi T\nonumber
\\
&+&m^2\int d^D x\sqrt{-g}\left( -\frac{1}{2} F_{\mu\nu}F^{\mu\nu}+\frac{4\Lambda}{D-2} A_\mu A^\mu\right)-2m^2\left(h_{\mu\nu}\nabla^\mu A^\nu-h\nabla_\mu A^\mu\right)\label{fullaction}
\\
&+&\left(2\Lambda-m^2\left(D-1\right)\right)\left( \frac{2m^2}{D-2}\nabla_\mu\phi\nabla^\mu\phi-\frac{D}{2} \frac{4m^4}{(D-2)^2}\phi^2 \right)\nonumber
\\
&+&\left.\left(-2m^2\right)\nabla_\mu A^\mu \phi\frac{2}{D-2}\left(2\Lambda - (D-1)m^2\right)+\frac{2m^2}{D-2}h\phi\left(-2\Lambda+m^2\left(D-1\right)\right)\right]\nonumber
\end{eqnarray}
It can be seen from Eq. (\ref{phikin}) that a specific value of the mass, $m^2=\frac{2\Lambda}{D-1}$ ($m^2=\frac{2\Lambda}{3}$ in 4 dimensions), renders the scalar field $\phi$ nondynamical. Moreover, we can see from Eq. (\ref{phiAmix}), (\ref{phihmix}), (\ref{phikin}) (or collectively from Eq. (\ref{fullaction})) that this value of mass also removes every existences of the field $\phi$. Thus, the massive gravity satisfying this special value of mass $m^2=\frac{2\Lambda}{D-1}$ is known as a partially-massless gravity.

\subsection{Equation of motion for the gravitational fields and their mode decomposition}
The partially massless graviton equation of motion is given by

\begin{itemize}
\item ${\rm\bf EOM\ for \ }\phi$:
\begin{equation}
0=\nabla_\mu\nabla_\nu T^{\mu\nu}+\frac{2\Lambda}{(D-2)(D-1)}T
\end{equation}
\item ${\rm\bf EOM\ for \ }A_{\mu} $:
\begin{equation}
\label{a-equation}
0=m^2(\nabla^\mu F_{\mu\nu} + \frac{4\Lambda}{D-2} A_\nu + \nabla^\mu h_{\mu\nu}-\nabla_\nu h)-\nabla^\mu T_{\mu\nu}
\end{equation}
\item ${\rm\bf EOM\ for \ }h_{\mu\nu} $:
\begin{eqnarray}
\label{h-equation}
\nonumber
0&=&\nabla^2 h_{\mu\nu}+\frac{2D}{(D-2)(D-1)}\Lambda h_{\mu\nu}+\nabla_\mu\nabla_\nu h -\nabla_{\lambda} \nabla_{\mu} h_{\nu}^{\ \lambda}-\nabla_{\lambda}\nabla_{\nu} h_{\mu}^{\ \lambda}
\\
&+&g_{\mu\nu}\left( \nabla_\rho\nabla_\lambda h^{\rho\lambda}-\nabla^2 h -\frac{2\Lambda}{(D-2)(D-1)}h\right)\nonumber
 \\ 
&-&\frac{2\Lambda}{D-1}(\nabla_\mu A_\nu+\nabla_\nu A_\mu-2\nabla_\rho A^\rho g_{\mu\nu})+T_{\mu\nu}.
\end{eqnarray}
\end{itemize}
When we operate $\nabla^\nu$ on the both sides of the Eq. (\ref{a-equation}), we get
\begin{equation}
\nabla^\alpha\left( A_\alpha + \frac{D-2}{4\Lambda}(\nabla^\nu h_{\nu\alpha}-\nabla_\alpha h) \right)=-\frac{T}{4\Lambda}.
\end{equation}
It can be easily shown that the trace of Eq. (\ref{h-equation}) is the same with this equation. Acting $\nabla^{\mu}$ on the both sides of Eq. (\ref{h-equation}) reproduces Eq. (\ref{a-equation}). To show this, we utilize the following identities:

\begin{eqnarray}
\nonumber
\nabla^\mu(\nabla^2h_{\mu\nu}-\nabla_\mu\nabla_\lambda h_\nu^{\ \lambda})&=&-R^{\ \beta\mu}
\nabla_\beta h_{\mu\nu}+R^{\beta\alpha}\nabla_\alpha h_{\beta\nu}+R_\nu^{\ \beta\mu\alpha}\nabla_\alpha h_{\mu\beta}\\ \nonumber
&&+\nabla^\alpha(R^\beta_{\ \alpha}h_{\beta\nu}-R_{\nu \ \alpha}^{\ \beta\ \mu}h_{\mu\beta}) \\ 
&\Rightarrow&\frac{2\Lambda}{(D-2)(D-1)}((D+1)\nabla^\mu h_{\mu\nu}-2\nabla_\nu h), \\
\nabla^\mu(g_{\mu\nu}\nabla_\rho \nabla_\lambda h^{\rho\lambda}-\nabla_\nu\nabla_\lambda h_\mu^{\ \lambda})&=&-R_{\beta\nu}\nabla_\lambda h^{\beta\lambda}\Rightarrow -\frac{2\Lambda}{D-2}\nabla_\lambda h
_\nu^{\ \lambda}, \\
\nabla^\mu(\nabla_\mu \nabla_\nu h-g_{\mu\nu}\nabla^2 h)&=&R^{\beta}_{\ \nu}\nabla_{\beta}h \Rightarrow \frac{2\Lambda D}{D-2}\nabla_\nu h,
\end{eqnarray}
where the arrows means that we use the properties of de Sitter space.

The equations of motion and the action enjoy the following gauge symmetry:
\begin{eqnarray}
\delta h_{\mu\nu}&=&2\nabla_{(\mu} \xi_{\nu)}+\frac{4\Lambda}{3(D-2)} \lambda g_{\mu\nu} \nonumber\\ 
\delta A_\mu&=&-\xi_\mu+\nabla_\mu \lambda \\
\delta \phi&=&-\lambda,\nonumber
\end{eqnarray}
and by a redefinition of the gauge parameters as 
$\xi_\mu\equiv\bar\xi_\mu+\nabla_\mu\lambda$,
\begin{eqnarray}
\delta h_{\mu\nu}&=&2\nabla_{(\mu} \bar\xi_{\nu)}+2\left(\nabla_\mu\nabla_\nu \lambda+\frac{2\Lambda}{3(D-2)} g_{\mu\nu}\lambda \right)\nonumber\\ 
\delta A_\mu&=&-\bar\xi_\mu \\
\delta \phi&=&-\lambda.\nonumber
\end{eqnarray}
Up to this point, we have seen the equations of motion and the gauge structures of this theory in covariant forms (in $D$ dimensions). To explicitly realize how the gauge symmetry projects out some degrees of freedom, leaving only dynamical degrees of freedom, we may consider this theory in 4 dimensions in the next section.

\subsection{Calculation using conformal transformation and $3+1$ decomposition}
Before explicitly realizing how the gauge symmetry projects out the nonphysical degrees of freedom, we can first consider the PM gravity from another approach. Since in this study we chose a specific coordinate to represent the de Sitter metric so that it appears to be conformal to the Minkowski metric, we can alternatively consider the PM action by the conformal transformation which can be found in general GR textbooks. We first consider an Einstein-Hilbert action with cosmological constant and the Fierz-Pauli mass term on a generic spacetime denoted by $\tilde{g}_{\mu\nu}$,
\begin{eqnarray}
S=\int d^4x \sqrt{-\tilde{g}} \left(\tilde{R}-2\Lambda\right) -\frac{1}{4}\sqrt{-\tilde g^0}m^2 \tilde{g}^{0\;\mu\alpha} \tilde{g}^{0\;\nu\beta}\left(\tilde{h}_{\mu\nu}\tilde h_{\alpha\beta}-\tilde h_{\mu\alpha} \tilde h_{\nu\beta}\right),\label{curvedaction}
\end{eqnarray}
where $\tilde{R}$ is a Ricci scalar corresponding to $\tilde{g}_{\mu\nu}$, the mass term is defined on a background $\tilde{g}^0_{\mu\nu}$, and the tensor field $\tilde{h}_{\mu\nu}\equiv\tilde{g}_{\mu\nu}-\tilde{g}^0_{\mu\nu}$.
In the case that $\tilde{g}_{\mu\nu}$ is related to $\mathcal{G}_{\mu\nu}$ by a conformal transformation, i.e. $\tilde{g}_{\mu\nu}=\omega^2 \mathcal{G}_{\mu\nu}$, then we have the following identity for the Ricci scalar, $\tilde{R}$, in generic $D$ dimensions,
\begin{eqnarray}
\tilde{R} &=& \omega^{-2}\mathcal{R}-2\left(D-1\right)\mathcal{G}^{\alpha\beta}\omega^{-3}\nabla_\alpha \nabla_\beta\,\omega-\left(D-1\right)\left(D-4\right)\mathcal{G}^{\alpha\beta}\omega^{-4} \nabla_\alpha\,\omega \nabla_\beta\,\omega,
\\
&=&\omega^{-2}\mathcal{R}-6\mathcal{G}^{\alpha\beta}\omega^{-3}\nabla_\alpha \nabla_\beta\,\omega \qquad\qquad \text{in $4D$ $(D=4)$},
\end{eqnarray}
where $\mathcal{R}$ is a Ricci scalar corresponding to $\mathcal{G}_{\mu\nu}$ and the covariant derivative $\nabla_\mu$ is evaluated on $\mathcal{G}_{\mu\nu}$.
Since we have chosen the de Sitter background in the following form,
\begin{equation}
ds^2=\frac{3}{\Lambda t^2}(-dt^2+d\vec x^2_3),
\end{equation}
then in this case $\omega^2=\frac{3}{\Lambda t^2}$. Moreover, we define $\tilde{g}^0_{\mu\nu}\equiv\omega^2\eta_{\mu\nu}$ and $\tilde h_{\mu\nu}\equiv\omega^2\mathcal{H}_{\mu\nu}$ (so that $\mathcal{G}_{\mu\nu}=\eta_{\mu\nu}+\mathcal{H}_{\mu\nu}$). Thus, according to the above identity and definitions we can express the action in Eq. (\ref{curvedaction}) as follows,
\begin{eqnarray}
S &=&\int d^4x \sqrt{-\tilde{g}}\left(\tilde{R}-2\Lambda\right)-\frac{1}{4}\sqrt{-\tilde g^0}m^2 \tilde{g}^{0\;\mu\alpha} \tilde{g}^{0\;\nu\beta}\left(\tilde{h}_{\mu\nu}\tilde h_{\alpha\beta}-\tilde h_{\mu\alpha} \tilde h_{\nu\beta}\right),\nonumber
\\
&=& \int d^4x \;\omega^4\sqrt{-\mathcal{G}}\left(\omega^{-2}\mathcal{R}-6\mathcal{G}^{\alpha\beta}\omega^{-3}\nabla_\alpha \nabla_\beta\,\omega-2\Lambda\right)\nonumber
\\
&&\quad-\frac{1}{4}m^2\omega^4 {\eta}^{\mu\alpha} {\eta}^{\nu\beta}\left(\mathcal{H}_{\mu\nu} \mathcal{H}_{\alpha\beta}-\mathcal{H}_{\mu\alpha} \mathcal{H}_{\nu\beta}\right),\nonumber
\\
&=& \int d^4x \;\omega^2\left[-\frac{1}{4}\partial^\lambda \mathcal{H}^{\mu\nu} \partial_\lambda \mathcal{H}_{\mu\nu} +\frac{1}{2}\partial^\mu \mathcal{H}^{\rho\alpha}\partial_\rho \mathcal{H}_{\mu\alpha}-\frac{1}{2}\partial^\mu \mathcal{H} \partial^\nu \mathcal{H}_{\mu\nu}+\frac{1}{4}\partial^\mu \mathcal{H}\partial_\mu \mathcal{H}\right]\nonumber
\\
&&\quad-\partial_\mu\partial_\nu \omega^2 \mathcal{H}^{\nu\lambda}{\mathcal{H}_\lambda}^\mu -\partial_\rho \omega^2 \mathcal{H}^{\rho\nu} \partial_\nu \mathcal{H} -\frac{1}{2}\partial^\mu\omega^2 \mathcal{H}\partial^\nu \mathcal{H}_{\mu\nu} +\frac{1}{2}\partial^i\partial_i\omega^2\left(\mathcal{H}^{\rho\lambda}\mathcal{H}_{\rho\lambda} -\frac{1}{2}\mathcal{H}^2\right)\nonumber
\\
&&\quad -6\left(\left(-\frac{1}{8}\mathcal{H}^2+\frac{1}{4}\mathcal{H}^{\mu\nu}\mathcal{H}_{\mu\nu}\right)\partial^\rho\omega\partial_\rho\omega +\frac{1}{2}\partial_\alpha\omega\partial_\rho\omega \mathcal{H}\mathcal{H}^{\alpha\rho} -\partial_\rho\omega\partial_\alpha\omega \mathcal{H}^{\rho\lambda}{\mathcal{H}_\lambda}^\alpha\right)\nonumber
\\
&&\quad -2\Lambda\omega^4 \left(\frac{1}{8}\mathcal{H}^2-\frac{1}{4}\mathcal{H}^{\mu\nu}\mathcal{H}_{\mu\nu}\right)-\frac{1}{4}m^2\omega^4\left(\mathcal{H}_{\mu\nu}\mathcal{H}^{\mu\nu}-\mathcal{H}^2\right),
\end{eqnarray}
where the upper indices in the last equality are defined on the Minkowski metric $\eta_{\mu\nu}$ and $\mathcal{H}=\eta^{\mu\nu}\mathcal{H}_{\mu\nu}$.

The metric enjoys global $SO(3)$ isometry. Therefore, we sort the gravitational modes by such symmetry.
Through the $3+1$ decomposition previously defined in Eq. (\ref{3+1decom1})-(\ref{3+1decom3}), the action can be expressed as
\begin{eqnarray}
S&=&\int d^4x \frac{3}{\Lambda t^2}\left(\frac{1}{4}\left(\dot{h}^{TT}_{ij}\dot{h}^{TT}_{ij}+\frac{1}{2}\partial_i\dot{E}^T_j \partial_i\dot{E}^{T}_j+24\dot{\Psi}^2 +\partial_i\partial_j\dot{E}\partial_i\partial_j \dot{E} -\frac{1}{3}\partial_i\partial_i \dot{E}\partial_j\partial_j\dot{E}\right)\right.\nonumber
\\
&&\quad+\frac{1}{2}\left(\partial_i B^{T}_j\partial_i B^T_j+\partial_i\partial_j B\partial_i\partial_j B\right)-\frac{1}{4}\left(\partial_ih^{TT}_{jk}\partial_i h^{TT}_{jk}+\frac{1}{2}\partial_k\partial_iE^T_j \partial_k\partial_i E^{T}_{j} \right.\nonumber
\\
&&\quad \left.+24\partial_i\Psi\partial_i\Psi +\partial_k\partial_i\partial_j E\partial_k\partial_i\partial_j E -\frac{1}{3}\partial_i\partial_j\partial_j E\partial_i\partial_k\partial_k E\right) +\partial_i\partial_i \dot{B}\Phi -\dot{\Phi}\partial_i\partial_iB\nonumber
\\
&&\quad+\left(\frac{1}{2}\partial_i\partial_i E^{T}_{j}B^T_j+2\partial_j\Psi\partial_jB+\frac{2}{3}\partial_j\partial_i\partial_iE\partial_jB\right)-\frac{1}{2}\partial_i\partial_iB\partial_j\partial_jB\nonumber
\\
&&\quad +\frac{1}{2}\left(\frac{1}{4}\partial_i\partial_iE^{T}_{k}\partial_j\partial_j E^T_k+4\partial_k\Psi\partial_k\Psi+\frac{8}{3}\partial_k\Psi\partial_k\partial_i\partial_iE+\frac{4}{3}\partial_k\partial_i\partial_iE\partial_k\partial_i\partial_iE\right) +3\dot{\Psi}\partial_i\partial_iB\nonumber
\\
&&\quad \left.-3\Psi\partial_i\partial_i\dot{B} + \left(\Phi+3\Psi\right)\left(2\partial_i\partial_i\Psi+\frac{2}{3}\partial_i\partial_i\partial_k\partial_kE\right)-9\dot{\Psi}^2 +6\partial_i\Phi\partial_i\Psi+9\partial_i\Psi\partial_i\Psi\right)\nonumber
\\
&&\quad -\frac{18}{\Lambda t^4}\Phi^2-\frac{36}{\Lambda t^3}\Phi\dot{\Psi}-\frac{9}{\Lambda t^3}\left(2\Phi+6\Psi\right)\partial_i\partial_iB \nonumber
\\
&&\quad-\frac{9}{4\Lambda^2t^4}m^2\left(-2\left(B^T_iB^{T}_{i}+\partial_iB\partial_iB\right) +h^{TT}_{ij}h^{TT}_{ij}+\frac{1}{2}\partial_iE^T_j\partial_iE^{T}_{j}+24\Psi^2 +\partial_i\partial_jE\partial_i\partial_jE\right.\nonumber
\\
&&\quad\left.-\frac{1}{3}\partial_i\partial_iE\partial_j\partial_jE-\left(24\Phi\Psi+36\Psi^2\right)\right). \label{fullaction2}
\end{eqnarray}
Thus, the action for tensor mode Eq. (\ref{tensoraction}) and vector mode Eq. (\ref{vectoraction}) can be read from the full action in Eq. (\ref{fullaction2}). In addition, from Eq. (\ref{fullaction2}), we can also show that all of the scalars defined in Eq. (\ref{3+1decom1})-(\ref{3+1decom3}) also do not contribute as dynamical degrees of freedom.

\subsection{Scalar mode action}
In addition to the tensor mode and the vector mode considered previously in the main article, we could also study the action for the scalar mode, though it does not contribute when the PM condition, $m^2=\frac{2\Lambda}{3}$ in $4D$, is met. From Eq. (\ref{fullaction2}) the scalar mode action can be read as follows,
\begin{eqnarray}
S_S &=& \int d^4 x \left(-\frac{18}{\Lambda t^2}\dot{\Psi}^2+\frac{1}{2\Lambda t^2}\partial^2 \dot{E}\partial^2 \dot{E}-\frac{6}{\Lambda t^2}\Psi\partial^2\Psi+\frac{1}{6\Lambda t^2}\partial_i\partial^2 E\partial_i\partial^2 E +\frac{12}{\Lambda t^2}\dot{\Psi}\partial^2 B \right.\nonumber
\\
&&\quad-\frac{2}{\Lambda t^2}\partial^2 \dot{E}\partial^2 B + \frac{2}{\Lambda t^2}\partial^2 \Psi \partial^2 E -\frac{12}{\Lambda t^2}\Phi \partial^2 \Psi +\frac{2}{\Lambda t^2}\Phi\partial^4 E +\frac{12}{\Lambda t^3}\Phi \partial^2 B -\frac{18}{\Lambda t^4}\Phi^2 \nonumber
\\
&&\quad \left.-\frac{36}{\Lambda t^3}\Phi \dot{\Psi} -\frac{9}{4\Lambda^2 t^4}m^2\left(-2\partial_i B\partial_i B -24\Psi^2 +\frac{2}{3}\partial^2 E\partial^2 E -24\Phi\Psi\right)\right),\label{scalaraction}
\end{eqnarray}
where $\partial^2\equiv\partial_i\partial_i$. 
There are four equations of motion as follows,
\begin{eqnarray}
\Phi &\to& -\frac{12}{\Lambda t^2}\partial^2\Psi +\frac{2}{\Lambda t^2}\partial^4E +\frac{12}{\Lambda t^3}\partial^2 B -\frac{36}{\Lambda t^4}\Phi - \frac{36}{\Lambda t^3}\dot{\Psi} +\frac{54}{\Lambda^2 t^4}m^2 \Psi =0, \label{Phieom}
\\
B &\to& \frac{12}{\Lambda t^2}\partial^2 \dot{\Psi} -\frac{2}{\Lambda t^2}\partial^4 \dot{E} + \frac{12}{\Lambda t^3}\partial^2 \Phi -\frac{9}{\Lambda^2 t^4}m^2 \partial^2 B=0,\label{Beom}
\\
E &\to& \partial_t\left(-\frac{1}{\Lambda t^2}\partial^4 \dot{E}\right) -\frac{1}{3\Lambda t^2}\partial^6 E +\partial_t\left(\frac{2}{\Lambda t^2}\partial^4 B\right) +\frac{2}{\Lambda t^2}\partial^4 \Psi +\frac{2}{\Lambda t^2}\partial^4 \Phi \nonumber
\\
&&-\frac{3}{\Lambda^2 t^4}m^2\partial^4 E=0,\label{Eeom}
\\
\Psi &\to& \partial_t\left(\frac{36}{\Lambda t^2}\dot{\Psi}\right) -\frac{12}{\Lambda t^2}\partial^2 \Psi -\partial_t\left(\frac{12}{\Lambda t^2}\partial^2 B\right) +\frac{2}{\Lambda t^2}\partial^4 E-\frac{12}{\Lambda t^2}\partial^2 \Phi +\partial_t\left(\frac{36}{\Lambda t^3}\Phi\right) \nonumber
\\
&&+\frac{9}{\Lambda^2 t^4}m^2 \left(12\Psi+6\Phi\right)=0.\label{Psieom}
\end{eqnarray}
In this context, it might not seem obvious that under the PM condition, $m^2=\frac{2\Lambda}{3}$ in $4D$,all of the scalar modes decouple from the action in Eq. (\ref{fullaction2}). To see it explicitly, we consider these equations of motion in a momentum space in the next section.
%
%
%
%

\subsubsection{Absence of the scalar modes}
One may see from the action in Eq. (\ref{fullaction2}) that some of the scalar modes do not propagate because of the absence of their kinetic terms. To see this explicitly, we can consider the scalar equations of motion. In a momentum space, Eq. (\ref{Phieom}) can be rewritten as
\begin{eqnarray}
\Phi = \frac{t^2}{3}k^2\Psi +\frac{t^2}{18}k^4E -\frac{t}{3}k^2 B -t\dot{\Psi}+\frac{3}{2\Lambda}m^2\Psi. \label{Phieom2}
\end{eqnarray}
Plugging Eq. (\ref{Phieom2}) into Eq. (\ref{Beom}) yields
\begin{eqnarray}
B=\frac{1}{9m^2+4\Lambda t^2 k^2}\left(2\Lambda t^2 k^2 \dot{E} +\frac{2}{3}\Lambda t^3 k^4 E +\left(4\Lambda t^3 k^2+18tm^2\right)\Psi\right).\label{Beom2}
\end{eqnarray}
Using Eq. (\ref{Phieom2}) and Eq. (\ref{Beom2}), Eq. (\ref{Psieom}) can be rewritten into
\begin{eqnarray}
\Psi = \frac{6k^4\Lambda^2 t^3 \dot{E}-2k^6\Lambda^2 t^4 E - 9m^2k^4t^2\Lambda E}{3A}, \label{Psieom2}
\end{eqnarray}
where
\begin{eqnarray}
A\equiv 3\left(9m^2+4\Lambda t^2 k^2\right)\left(3m^2-2\Lambda\right)+4k^4\Lambda^2 t^4.
\end{eqnarray}
Eventually, by using Eq. (\ref{Phieom2}), (\ref{Beom2}), (\ref{Psieom2}), we obtain the equation of motion for $E$ as follows,
\begin{flalign}
&0=m^2k^4\left(2\Lambda-3m^2\right)\times&\nonumber
\\
&\left[\partial_t\left(\frac{\dot{E}}{\Lambda t^2A}\right)+\frac{9m^2\left(9m^2+7\Lambda t^2 k^2\right)\left(3m^2-2\Lambda\right)+4\Lambda^2 t^4 k^4 \left(12m^2-2\Lambda+\Lambda t^2 k^2\right)}{\Lambda^2 t^4 A^2}E\right]\label{Eeom}
\end{flalign}
In generic massive gravity theory, $E$ propagates as the fifth degree of freedom. On the $4D$ de Sitter space, however, $E$ can be made to decouple from the massive gravity by choosing the PM condition, $m^2=\frac{2\Lambda}{3}$. 
From Eq. (\ref{Eeom}), this is obvious that the scalar equation of motion vanishes when the PM condition is satisfied.

It is also worthwhile to note that, from the action in Eq. (\ref{scalaraction}), it looks like the $\Psi$ mode should propagate as another ghostly degree of freedom of the theory since $\Psi$ appears in the action as $-\dot{\Psi}^2$. However, the reason behind the absence of $\Psi$ mode is due to the specific form of the mass term $-m^2\left(h^{\mu\nu}h_{\mu\nu}-h^2\right)$, also known as the Fierz-Pauli tuning, which kills this unwanted degree of freedom  (see Ref. \cite{Hinterbichler:2011tt} for a good review of the Fierz-Pauli massive gravity).


\begin{thebibliography}{9}
\bibitem{VanNieuwenhuizen:1973fi} 
  P.~Van Nieuwenhuizen,
  Nucl.\ Phys.\ B {\bf 60}, 478 (1973).
  doi:10.1016/0550-3213(73)90194-6

\bibitem{Deser1}
S. Deser and C. Teitelboim, Phys. Rev. D 13, 1592(1976), S. Deser, J. Phys. A
15, 1053 (1982).

\bibitem{Moon:2014gaa} 
  S.~Moon, S.~J.~Lee, J.~Lee and J.~H.~Oh,
  J.\ Korean Phys.\ Soc.\  {\bf 67}, no. 3, 427 (2015)
  doi:10.3938/jkps.67.427
  [arXiv:1405.4934 [hep-th]].

\bibitem{Lee:2018bud} 
  H.~Lee, S.~Han, H.~Yoon, J.~Kim and J.~H.~Oh,
  arXiv:1801.08665 [hep-th].

\bibitem{Deser:2006zx} 
  S.~Deser and A.~Waldron,
  Phys.\ Rev.\ D {\bf 74}, 084036 (2006)
  doi:10.1103/PhysRevD.74.084036
  [hep-th/0609113].

\bibitem{deRham:2013wv} 
  C.~de Rham, K.~Hinterbichler, R.~A.~Rosen and A.~J.~Tolley,
  Phys.\ Rev.\ D {\bf 88}, no. 2, 024003 (2013)
  doi:10.1103/PhysRevD.88.024003
  [arXiv:1302.0025 [hep-th]].

\bibitem{Deser:2013uy} 
  S.~Deser, M.~Sandora and A.~Waldron,
  Phys.\ Rev.\ D {\bf 87}, no. 10, 101501 (2013)
  doi:10.1103/PhysRevD.87.101501
  [arXiv:1301.5621 [hep-th]].

\bibitem{Hinterbichler:2014xga} 
  K.~Hinterbichler,
  Phys.\ Rev.\ D {\bf 91}, no. 2, 026008 (2015)
  doi:10.1103/PhysRevD.91.026008
  [arXiv:1409.3565 [hep-th]].

\bibitem{Fierz:1939ix} 
  M.~Fierz and W.~Pauli,
  Proc.\ Roy.\ Soc.\ Lond.\ A {\bf 173}, 211 (1939).
  doi:10.1098/rspa.1939.0140

\bibitem{Chen:2012au} 
  T.~j.~Chen, M.~Fasiello, E.~A.~Lim and A.~J.~Tolley,
  JCAP {\bf 1302}, 042 (2013)
  doi:10.1088/1475-7516/2013/02/042
  [arXiv:1209.0583 [hep-th]].

\bibitem{Gambini:1980wm}
  R.~Gambini and A.~Trias,
  Phys.\ Rev.\ D {\bf 22} (1980) 1380.
  doi:10.1103/PhysRevD.22.1380

\bibitem{Veltman:1975vx} 
  M.~J.~G.~Veltman,
  Conf.\ Proc.\ C {\bf 7507281}, 265 (1975).

\bibitem{Christensen:1979iy} 
  S.~M.~Christensen and M.~J.~Duff,
  Nucl.\ Phys.\ B {\bf 170}, 480 (1980).
  doi:10.1016/0550-3213(80)90423-X

\bibitem{Allen:1986tt} 
  B.~Allen and M.~Turyn,
  Nucl.\ Phys.\ B {\bf 292}, 813 (1987).
  doi:10.1016/0550-3213(87)90672-9


\bibitem{Deser:2001xr} 
  S.~Deser and A.~Waldron,
  Phys.\ Lett.\ B {\bf 513}, 137 (2001)
  doi:10.1016/S0370-2693(01)00756-0
  [hep-th/0105181].
  
\bibitem{Hinterbichler:2011tt} 
  K.~Hinterbichler,
  Rev.\ Mod.\ Phys.\  {\bf 84}, 671 (2012)
  doi:10.1103/RevModPhys.84.671
  [arXiv:1105.3735 [hep-th]].

\bibitem{Boulanger:2020bah}
N.~Boulanger, S.~Garcia-Saenz and L.~Traina,
Phys. Part. Nucl. Lett. \textbf{17}, no.5, 687-691 (2020)
doi:10.1134/S1547477120050064

\bibitem{Garcia-Saenz:2014cwa}
S.~Garcia-Saenz and R.~A.~Rosen,
JHEP \textbf{05}, 042 (2015)
doi:10.1007/JHEP05(2015)042
[arXiv:1410.8734 [hep-th]].

\bibitem{Joung:2014aba}
E.~Joung, W.~Li and M.~Taronna,
Phys. Rev. Lett. \textbf{113}, 091101 (2014)
doi:10.1103/PhysRevLett.113.091101
[arXiv:1406.2335 [hep-th]].

\bibitem{Apolo:2016ort}
L.~Apolo and S.~F.~Hassan,
Class. Quant. Grav. \textbf{34}, no.10, 105005 (2017)
doi:10.1088/1361-6382/aa69f7
[arXiv:1609.09514 [hep-th]].

\bibitem{Garcia-Saenz:2015mqi}
S.~Garcia-Saenz, K.~Hinterbichler, A.~Joyce, E.~Mitsou and R.~A.~Rosen,
JHEP \textbf{02}, 043 (2016)
doi:10.1007/JHEP02(2016)043
[arXiv:1511.03270 [hep-th]].



\bibitem{Garcia-Saenz:2018wnw}
S.~Garcia-Saenz, K.~Hinterbichler and R.~A.~Rosen,
JHEP \textbf{11}, 166 (2018)
doi:10.1007/JHEP11(2018)166
[arXiv:1810.01881 [hep-th]].

\bibitem{Bernard:2017tcg}
L.~Bernard, C.~Deffayet, K.~Hinterbichler and M.~von Strauss,
Phys. Rev. D \textbf{95}, no.12, 124036 (2017)
[erratum: Phys. Rev. D \textbf{98}, no.6, 069902 (2018)]
doi:10.1103/PhysRevD.95.124036
[arXiv:1703.02538 [hep-th]].

\bibitem{Hinterbichler:2015nua}
K.~Hinterbichler and R.~A.~Rosen,
Phys. Rev. D \textbf{92}, no.10, 105019 (2015)
doi:10.1103/PhysRevD.92.105019
[arXiv:1507.00355 [hep-th]].

\bibitem{Enander:2015dpn}
J.~Enander,``Cosmic tests of massive gravity,'' Thesis.

\bibitem{Alexandrov:2014oda}
S.~Alexandrov and C.~Deffayet,
JCAP \textbf{03}, 043 (2015)
doi:10.1088/1475-7516/2015/03/043
[arXiv:1410.2897 [hep-th]].

\bibitem{Deser:2014fta}
S.~Deser, K.~Izumi, Y.~C.~Ong and A.~Waldron,
Mod. Phys. Lett. A \textbf{30}, 1540006 (2015)
doi:10.1142/S0217732315400064
[arXiv:1410.2289 [hep-th]].

\bibitem{Brito:2013yxa}
R.~Brito, V.~Cardoso and P.~Pani,
Phys. Rev. D \textbf{87}, no.12, 124024 (2013)
doi:10.1103/PhysRevD.87.124024
[arXiv:1306.0908 [gr-qc]].

\end{thebibliography}
\end{document}